\documentclass[12pt, letterpaper]{article}

\usepackage{amsmath, amssymb, graphics, epsfig, graphicx, color}
\usepackage{epsf}
\usepackage{epstopdf}

\newcommand{\nc}{\newcommand}
\nc{\ba}{\begin{eqnarray}}
\nc{\ea}{\end{eqnarray}}

\newcommand{\calR}{{\cal{R}}}

\nc{\im}{{ \mathrm{Im} } }

\begin{document}

\vspace{5mm}
\vspace{0.5cm}
\begin{center}

{\large \bf Primordial Statistical Anisotropies:\\  \vspace{0.3cm}The Effective Field Theory Approach
}
\\[0.5cm]

{ Ali Akbar Abolhasani$^{1}$, 
Mohammad Akhshik$^{1, 2}$,  Razieh Emami$^{3}$,   Hassan Firouzjahi$^2$}
 \vspace{0.3cm}

{\small \textit{$^1$Department of Physics, Sharif University of Technology, Tehran, Iran
}}\\
{\small \textit{$^2$School of Astronomy, Institute for Research in Fundamental Sciences (IPM) \\ P.~O.~Box 19395-5531, Tehran, Iran\\
}}
{\small \textit{$^3$ Institute for Advanced Study, The Hong Kong University
of Science and Technology, Clear Water Bay, Kowloon, Hong Kong
}}

\end{center}

\vspace{.8cm}

\hrule \vspace{0.3cm}

\begin{abstract}

In this work we present the effective field theory of  primordial statistical anisotropies generated during  anisotropic inflation involving a background $U(1)$ gauge field.  Besides the usual 
Goldstone boson associated with the breaking of time diffeomorphism we have two additional Goldstone bosons associated with the breaking of spatial diffeomorphisms.  We further identify these two new Goldstone bosons with the expected two transverse degrees of the $U(1)$ gauge field fluctuations. Upon defining the appropriate unitary gauge, we present the most general quadratic action which respects the remnant symmetry in the unitary gauge. The  interactions between  various Goldstone bosons  leads to statistical anisotropy in curvature perturbation power spectrum. 
Calculating the general results for power spectrum anisotropy, we recover the previously known results in specific models  of anisotropic inflation. In addition, we present novel results for statistical anisotropy in models with non-trivial sound speed for inflaton fluctuations. Also we identify the interaction which leads to birefringence-like effects in anisotropic power spectrum in which the speed of gauge field fluctuations depends on the direction of the mode propagation and the two polarization of gauge field fluctuations contribute differently in statistical anisotropy. As another interesting application, our EFT approach naturally captures interactions generating  parity violating statistical anisotropies.

\end{abstract}

\newpage
\section{Introduction}

Inflation is widely accepted as the leading paradigm for early universe with its basics  predictions  being well consistent with cosmological observations  \cite{Ade:2013lta, Ade:2013uln}. During inflation,  the quantum fluctuations of inflaton field(s) 
and the metric  are amplified to cosmological scales which 
induce nearly scale-invariant, nearly adiabatic and nearly Gaussian perturbations on cosmic microwave background (CMB) maps and large scale structures which are in good agreements with data. The simplest models of inflation are based on a
scalar field which is minimally coupled to gravity and  rolls slowly over a near flat potential.

Despite  all the observational successes of inflation,  there is no fundamental understanding
of mechanism of inflation. For example, the fundamental questions such as  
what was the stage of universe prior to inflation or 
what is the nature of inflaton field are left unanswered within the current working paradigm of inflation. Lacking 
a fundamental understanding of the mechanism of inflation,  there are many phenomenological  
models of inflation which are consistent with data. Naturally one is lead to ask  how far one can capture the  most robust predictions of models of inflation without relying on particular realization of inflation model building.  Effective Field Theory (EFT) of inflation \cite{Cheung:2007st} has been a successful program to answer this question, for a review of general EFT methods see \cite{Manohar:1996cq, Burgess:2007pt}.  In the logic of EFT all interactions which are  compatible with the underlying symmetries should be considered.  Then depending on how one turn on particular interactions governing the dynamics of  the light field, different inflationary models are realized. 
EFT approach was particularly successful in models of single field inflation in 
classifying their predictions for power spectrum and bispectrum. Similarly, one can extend  the method of EFT of inflation 
to models of multiple fields  inflation \cite{Senatore:2010wk}.

Most of models of inflation are based on scalar field dynamics. This is mainly motivated from the fact that the 
scalar fields are by construction  spin-zero fields, naturally apt to generate isotropic cosmological backgrounds, 
a fundamental requirement of cosmological principle. Having this said, it is natural to examine the role of
other type of fields during inflation. In particular, vector fields and gauge fields are ubiquitous in Standard Model of
particle physics and in quantum field theory. Therefore, one expects that models of inflation with vector fields
can have interesting theoretical motivations which also can be directly confronted with the data. Anisotropic inflation 
is such a realization based on gauge field dynamics  which have captured 
significant interests in the literature. In most attractive realization of anisotropic inflation, a $U(1)$ gauge field is turned on at the background level with a non-zero electric field energy density.  In order to sustain the background electric field energy density and to endow a scale-invariant  spectrum for the  gauge field perturbations,  the gauge field is coupled to the inflaton field.   
Observationally models of anisotropic inflation predict statistical anisotropy in CMB map which
can be tested observationally. 

Here our goal is to extend the logic of EFT to the setup of anisotropic inflation which generate statistical anisotropies. 
We assume the matter sector contains a scalar field $\phi$, playing the role of inflaton, and a $U(1)$ gauge field within the Einstein gravity. With these minimal assumptions, we look for all possible interaction allowed by the underlying symmetries. This generality allows us to go beyond the  model-dependent picture 
of anisotropic inflation and to look for new types of interactions between the inflaton field and the gauge field perturbations. Consequently, we re-derive the previously known results of the power spectrum statistical anisotropies. In addition,  we obtain new results for power spectrum statistical anisotropies beyond the known results. 

The important starting point to construct the EFT of inflation is to identify the symmetries of the problem at hand.
In models of single field inflation, this task is well-understood. To start one chooses a time foliation, known as the unitary gauge,   
such that the scalar field remains homogeneous. Consequently, all perturbations are transferred into metric sector. In this view, the symmetry of the system contains all coordinate transformation which leaves the time foliation intact. In other words, the general four-dimensional  diffeomorphism invariance $x^\mu \rightarrow x^\mu + \xi^\mu$ is reduced to the three-dimensional transformation $x^i \rightarrow x^i + \xi^i( x^\nu)$. The building blocks of this remnant symmetry transformation in unitary gauge  are  $g^{00}$, $K_{ij}$ etc in which
the latter is the extrinsic curvature of the constant time hypersurface. Equipped with these building blocks  one writes down  all the possible interactions consistent with the remnant symmetry. Equivalently, 
 one can look at the same problem in an arbitrary coordinate system in which the time coordinate is not fixed.  Physically, this corresponds to restoring a scalar field degree of freedom, the so-called Goldstone boson $\pi$,  which captures the fluctuations of scalar field perturbations. The advantage
 of the EFT is when one goes to the decoupling limit in which one can neglect the gravitational back-reaction of
 $\pi$ with the metric perturbations, corresponding to $M_P \rightarrow \infty$, in which the fluctuations of $\pi$ capture
 the main results of the power spectrum and the bispectrum to leading order in terms of the slow roll parameters.  
 
Now in our setup of anisotropic inflation with an additional gauge field, the role of remnant symmetry and the choice of unitary gauge  is somewhat obscure. As in conventional case, we still choose the time foliation such that 
to keep the scalar field homogeneous, $\delta \phi=0$. As for the gauge field
excitations, we can define the unitary gauge to be the gauge in which  $\delta A_\mu=0$. However, this requirement is ambiguous as one has the $U(1)$ gauge symmetry 
$\delta A_\mu \rightarrow \delta A_\mu + \partial_\mu {\cal F} $  with ${\cal F} $ an arbitrary  scalar  in which  the gauge field fluctuations can
be turned on again. Therefore, an important task in defining our unitary gauge  is to properly take into account the role of $U(1)$ gauge transformation  along with space-time coordinate transformations to correctly identify  the remnant symmetry of the setup.

The paper is organized as follows. In Section \ref{Symmetry sec.} we identify the symmetries and the degrees of freedom
and present  the invariant action in unitary gauge. In Section \ref{Goldstone sec.} we restore
the Goldstone bosons and re-write the action, including the action of the free fields and the interactions, in terms of
the Goldstone bosons. In  Section \ref{Anisotropy sec.} we calculate the power spectrum anisotropy generated from various interactions followed by summary and discussions in Section \ref{Summary sec.}. We comment that this work is exclusively devoted to power spectrum analysis.


\section{Symmetries and Degrees of Freedom}
\label{Symmetry sec.}

In this Section we briefly review the setup of anisotropic inflation and then identify the physical degrees of freedom and the symmetries of the system to properly identify the starting unitary gauge. 

\subsection{Anisotropic Inflation }

As discussed before, our setup contains a scalar field $\phi$ playing the role of the inflaton field and a $U(1)$ gauge field.
The gauge field has a background value which without loss of generality can be taken to be along the x-direction so the gauge
field has the form $\overline A^\mu = (0,\overline{A}^1(t),0,0)$ in which an overline indicates the background quantity. This
also induces a background electric field energy density, breaking the isotropy so the background geometry is in the form
of Bianchi type I  universe. With this choice of the background gauge field, we still have the rotational symmetry in two-dimensional  $yz$ plane. 

As mentioned before, in usual models of anisotropic inflation in order for the background electric field to  survive  the dilution from  the exponential expansion,  the gauge field is coupled to the inflaton field in the form $f(\phi)^2 F_{\mu \nu} F^{\mu \nu}$. The functional form of   $f(\phi)$ is determined by the potential $V(\phi)$ but in terms of scale factor $a(t)$, one needs to choose $f(\phi) \propto a(t)^{-2}$ in order for the background electric field energy density to furnish a nearly constant and sub-leading portion of the total energy density. At the level of perturbations, this specific form of $f(\phi)$ helps to maintain a  scale invariant power spectrum for the gauge field fluctuations. For a review on anisotropic inflation see \cite{Emami:2015qjl} and for 
various works related to anisotropic  power spectrum and bispectrum see \cite{Watanabe:2009ct, Watanabe:2010fh, Soda1, Emami1, Emami:2013bk, Abolhasani:2013zya, Abolhasani:2013bpa,  Chen:2014eua, Bartolo:2012sd, Shiraishi:2013vja, Shiraishi:2013oqa,  various, Naruko:2014bxa, Kim:2013gka}. Also see \cite{various2} 
for different realizations of statistical anisotropies.  

The contribution of the gauge fields into curvature perturbation  power spectrum $P_{\calR}$
is in the form of quadrupole anisotropy parametrized as follows \cite{Ackerman:2007nb, Pullen:2007tu}
\ba
\label{g*}
P_{\calR}({\bf k}) =   P_{\calR}^{(0)} \left( 1 + g_* (\widehat {\bf n}\cdot  \widehat {\bf k})^2  \right) \, ,
\ea
in which $P_{\calR}^{(0)}$ is the leading isotropic power spectrum, 
${\bf k}$ is the mode of interest in Fourier space and $\widehat{\bf n}$ represents the direction of anisotropy. The parameter $g_{*}$ measures the strength of statistical anisotropy. Observational constraints from Planck data require
\cite{Kim:2013gka, Ade:2013uln} $| g_*| \lesssim 10^{-2}$. 

For broad class of potentials, it is shown in \cite{Watanabe:2009ct} that with the appropriate form of
the coupling $f(\phi)$, the system reaches the attractor regime in which the contribution from the electric field energy density reaches a constant and subdominant portion of the total energy density. Denoting the fraction of the electric energy density to total energy density by the parameter $R$, the correction in curvature perturbation 
power spectrum anisotropy in  models with simple chaotic potential  
is obtained to be 
\ba
g_* =   \left( \frac{48 R}{\epsilon} \right) N^2
\ea
in which  $\epsilon$ is the usual slow-roll parameter and $N$ represents the number of e-folds when the mode of interest 
$k$ leaves the horizon till the end of inflation. We mention that the $N^2$-dependence of power anisotropy is a generic feature  expected from the accumulative IR effects of the scale-invariant gauge field fluctuations \cite{Bartolo:2012sd}. Imposing the observational constraints on $g_*$ implies that $\frac{R}{\epsilon} \lesssim 10^{-5}$.
In addition, the bispectrum and the trispectrum analysis of the model were performed in \cite{Abolhasani:2013zya, Shiraishi:2013vja, Shiraishi:2013oqa} in which the amplitude of local-type non-Gaussianity is obtained to be 
$f_{NL} \sim g_* f_{NL}$ with non-trivial anisotropic shape of local-type non-Gaussianity. 

We mention that it is shown in \cite{Naruko:2014bxa} that for reasonable values of model parameters consistent with observations it is hard to reach the attractor regime ``during inflation". Nevertheless,  one can assume that the duration of inflation somewhat exceeds the minimal $60$ e-folds so the gauge filed settles down to its attractor solution well before the observable modes exit the horizon.


\subsection{Unitary Gauge and the General Action }

After briefly reviewing the models of anisotropic inflation, here we start our study of EFT for these setups. 

As in  \cite{Cheung:2007st} our starting job is to identify the physical degrees of freedom and 
to determine the proper unitary gauge. Following  the  logic of   EFT of  single field inflation 
\cite{Cheung:2007st}, in our setup 
one can define the unitary gauge as the gauge in which all matter perturbations  $\delta \phi$ and $\delta A^\mu$ are turned off so
\begin{equation}
\label{UG}
A^\mu =(0,\overline{A}^1(t),0,0), \qquad \phi = \overline{\phi}(t)  \quad \quad    \mathrm{ (unitary ~  gauge) } \, .
\end{equation}
Consequently, all perturbations are carried by the metric sector.

The  condition $\delta \phi=0$ can be satisfied easily as in \cite{Cheung:2007st} by appropriate foliation of space-time in which the surfaces of constant time  coincide with uniform $\phi$ surfaces. This is motivated from the fact that the  inflaton field $\phi(t)$ can be used as the physical clock. 

As for the gauge field the situation is more non-trivial. First 
note that we work  with the contravariant components $\delta A^\mu$ instead of the more natural covariant vector  $\delta A_\mu$ . The reason is that in fixing unitary gauge we have to choose our coordinate system such that all perturbations of the field vanish and all degrees of freedom appear in metric. As one may easily check, all covariant components of $\delta A_\mu$ are transformed by $\xi ^1$ and  therefore the condition $\delta A_\mu =0$ does not involve $\xi^\mu$ with
$\mu \neq 1$. However, as we shall see momentarily, $\delta A^\mu$ transformation 
is controlled by all components of $\xi ^\mu$ and consequently we may easily achieve $\delta A^\mu=0$ with the aid of a combination of coordinate transformation and U(1) gauge symmetry.

Second and more importantly, the condition $\delta A^\mu=0$ should be taken with care. It is true that part of spatial diffs can be fixed as the condition to put gauge field on its background value.  However, things become non-trivial if one keeps in mind that the system should also be invariant
under the $U(1)$ gauge transformation 
\ba
\label{U1-gauge}
A^\mu \rightarrow A^\mu + \nabla^\mu  {\cal F} = A^{\mu} + g^{\mu \nu} \partial_\nu {\cal F} 
\ea
in which ${\cal F} (x^\nu)$ is a scalar. Therefore, even if  we start with the unitary gauge $\delta A^\mu=0$, then the gauge field excitations can be restored by the $U(1)$ transformation Eq. (\ref{U1-gauge}).  As a result, in order to read off the correct physical
degrees of freedom of the gauge field one has to look into the transformation of the gauge field perturbations both under 
the $U(1)$ gauge  transformation (\ref{U1-gauge}) and also under the general coordinate transformation 
\ba
\label{coordinate}
x^\mu \rightarrow x'^\mu=   x^\mu + \xi^\mu(x^\nu) \, .
\ea 
Combining the transformations   (\ref{U1-gauge}) and (\ref{coordinate}) the gauge field perturbations transform effectively as 
\begin{equation}
\label{U1-coordinate}
\delta A^\mu \rightarrow  \delta A^\mu + \overline A^1\partial _1 \xi ^\mu + g^{\mu\alpha}\partial _\alpha {\cal F} .
\end{equation}
As usual, it is more convenient to decompose the four vector $\xi^\mu$ into the transverse and the longitudinal parts,  
$\xi^\mu_T$ and $\xi^\mu_L$ respectively as follows
\begin{equation}
\xi ^\mu = \nabla ^\mu \xi _L+\xi ^\mu _T=g^{\mu\alpha}\partial _\alpha \xi _L + \xi ^\mu _T \, ,
\end{equation}
subject to $\nabla _\mu \xi ^\mu _T=0$.   

Plugging these decompositions in transformation (\ref{U1-coordinate}) yields 
\ba
\delta A^\mu  &\rightarrow  &
 \delta A^\mu + \overline A^1\partial _1 \xi ^\mu _T+g^{\mu\alpha}\partial _\alpha {\cal F} + \overline A^1\partial _1 \left(g^{\mu\alpha}\partial _\alpha \xi _L\right) , \nonumber \\
&=& \delta A ^\mu + \overline A^1\partial _1 \xi ^\mu _T+ \overline A^1\left(\partial _1 g^{\mu \alpha}\right)\partial _\alpha \xi _L+
 \dot{ \overline  A}^1 g^{\mu 0}\partial _1 \xi _L + g^{\mu \alpha} \partial _\alpha \left( \overline A^1\partial _1 \xi _L + {\cal F} \right).
\ea
The above  transformation encodes both the $U(1)$ transformation (\ref{U1-gauge}) and the coordinate transformation 
(\ref{coordinate}). Now we are able to see  how the unitary gauge defined in  Eq. (\ref{UG}) is feasible. First, we note that 
by choosing ${\cal F} = -\overline A^1\partial _1 \xi _L$ we can always cancel the last term above.  In other words, with the aid of $U(1)$ symmetry we can partially cancel the variation  in $\delta A^\mu$ which is caused by $\xi _L$ coordinate transformation. 
Also, we have to remember that in unitary gauge, we already fixed $\xi^0$ to put inflaton on its background value. Now the unitary gauge defined in Eq. (\ref{UG}), with all matter perturbations turned off,  are subject to  remnant symmetry $x^\mu \rightarrow x^\mu +\xi ^\mu$ in which 
\begin{align}
\xi ^\mu _T=\xi ^\mu _T(t,y,z), \qquad \left(\partial _1 g^{\mu \alpha}\right) \partial _\alpha \xi _L=\frac{\dot{\overline A^1 }}{\overline A^1}g^{\mu 0}\partial _1 \xi _L,    \quad \quad \mathrm{ (remnant ~  symmetry) } \, 
\label{UG-symmetry}
\end{align}
excluding $\xi ^0$ component. It is curious that $\xi ^\mu _T$ is independent of the $x$ coordinate. 
Note that by  remnant symmetry we mean that every term in the EFT action should be  invariant under the above symmetries in unitary gauge. The above remnant
symmetry in our unitary gauge  should be compared with the remnant symmetry in
isotropic model containing only a single scalar field \cite{Cheung:2007st} in which $x^i \rightarrow x^i + \xi( t, x, y, z)$.

Having obtained the remnant symmetry of our system,  our next job is to construct all scalars which are invariant under these remnant symmetries. It is important to note that we have fixed the unitary gauge such that there is no matter field perturbations and
all perturbations are encoded in metric sector $\delta g_{\alpha \beta}$.  Consequently, all terms in the action in the unitary gauge  are constructed from the metric perturbations and their derivatives.
 
As in the  setup of EFT of \cite{Cheung:2007st} involving a single scalar field \cite{Cheung:2007st},  $g^{00}$ 
is a scalar so we keep $\delta g^{00}$ as one of our main building block  to write down the action in unitary gauge. 
As for other building blocks constructed from metric sector, we note that neither $\delta g^{11}$ nor any of other metric 
component transform as scalars under the remnant symmetry Eq.  \eqref{UG-symmetry} so we should look for
more non-trivial combinations. However, we see that our symmetry conditions in Eq. \eqref{UG-symmetry} involve
$\partial_1$ and $\partial_\alpha$. In particular, we note that  $\partial_1 \xi ^i_T=0 $. This suggests that if we works
with the metric perturbations with the lower indices, $\delta g_{\alpha \beta}$, we encounter the objects 
$\partial_\alpha \xi_L$ and $\partial_1 \xi_T$ which help to construct the desired scalars (or tensor). 
Since $\partial_1 \xi_T=0$, it seems that $g_{1 \alpha}$ may be a useful quantity to start with. However, under
the general coordinate transformation Eq. (\ref{coordinate}) we obtain 
\ba
g_{\alpha 1} &\rightarrow &\Lambda ^{\alpha ^\prime}_\alpha \Big(g_{\alpha ^\prime 1}+g_{\beta ^\prime \alpha ^\prime}\partial _1g^{\beta ^\prime \lambda}\partial _\lambda \xi _L\Big) \nonumber \\
&= &\Lambda ^{\alpha ^\prime}_\alpha \Big(g_{\alpha ^\prime 1}+\partial _1\partial _\alpha \xi _L +\frac{\dot{A}^1}{A^1}\delta ^0_{\alpha ^\prime}\partial _1 \xi _L\Big),
\ea
in which we have defined $\Lambda ^{\alpha ^\prime}_\alpha\equiv \frac{\partial x^{\alpha ^\prime}}{\partial x^\alpha}$. 
We note that the presence of last two terms involving $\xi_L$
tells us that $g_{1 \alpha}$ is not a  four-vector with respect to the free index $\alpha$. 
Similarly,  $g_{11}$ is not invariant under the remnant symmetry Eq.  \eqref{UG-symmetry},  
so unlike $g^{00}$, $g_{11}$  can not be used as a starting building block as expected from the above discussions. 
This indicates that
we have to use a more nontrivial combination of  $g_{1 \alpha}$ and its derivatives to construct the proper scalar, 
 four-vector or four-tensor.  
 
Now looking at the derivative of $g_{\alpha 1}$ we obtain 
\begin{align}
\partial _\beta g_{\alpha _1}&\rightarrow \Lambda ^{\beta ^\prime}_ \beta\Lambda ^{\alpha ^\prime}_\alpha \Bigg[ \partial _{\beta ^\prime}g_{\alpha ^\prime 1}+g_{\lambda 1}\xi ^{\lambda}_{,\alpha ^\prime \beta ^\prime}+\partial _1 \partial _{\alpha ^\prime}\partial _{\beta ^\prime} \xi _L + \partial _{\beta ^\prime}\Big(\frac{\overline{\dot{A}^1}}{\overline {A^1}}\partial _1 \xi _L\Big) \delta ^0_{\alpha ^\prime}\Bigg]  \, .
\end{align}
As before, the presence of $\xi_L$ and the second term in the bracket, prevent $\partial _\beta g_{\alpha _1}$ to be 
a four-tensor. However, we note that upon anti-symmetrization with respect to $\alpha $ and $\beta$ the second and the third terms  in the bracket above cancel and we obtain  
\begin{equation}
\label{transform-G-1}
\partial _\beta g_{\alpha _1}-\partial _\alpha g_{\beta 1}\rightarrow \Lambda ^{\beta ^\prime}_ \beta\Lambda ^{\alpha ^\prime}_\alpha \Bigg[ \partial _{\beta ^\prime}g_{\alpha ^\prime 1}-\partial _{\alpha ^\prime}g_{\beta ^\prime 1}+ \partial _{\beta ^\prime}\Big(\frac{\overline{\dot{A}^1}}{\overline {A^1}}\partial _1 \xi _L\Big) \delta ^0_{\alpha ^\prime}-\partial _{\alpha ^\prime}\Big(\frac{\overline{\dot{A}^1}}{\overline {A^1}}\partial _1 \xi _L\Big) \delta ^0_{\beta ^\prime}\Bigg] 
\end{equation}
Only if we can get rid of the term containing $\xi_L$ above, we can obtain a four-tensor. For this purpose, consider the
transformation of the following combination
\begin{align}
g_{\beta 1} \delta ^0_\alpha - g_{\alpha 1}\delta ^0_\beta &\rightarrow \Lambda ^{\beta ^\prime}_\beta\Lambda ^{\alpha ^\prime} _\alpha \bigg( g_{\beta ^\prime 1}\delta ^0_{\alpha ^\prime} - g_{\alpha ^\prime 1}\delta ^0_{\beta ^\prime}+\partial _1 \partial _{\beta ^\prime} \xi _L \delta ^0_{\alpha ^\prime}-\partial _1 \partial _{\alpha ^\prime}\delta ^0_{\beta ^\prime}\bigg) \, .
\end{align}
Combining the above equation with Eq. (\ref{transform-G-1}) we  are able to cancel out the undesired term 
in Eq. (\ref{transform-G-1}) containing $\xi_L$ with  the right combination of the above term. Hence if we define the quantity $G_{\alpha \beta}$ via
\begin{equation}
G_{\alpha \beta} \equiv \partial _\alpha  g_{\beta 1}-\partial _\beta g_{\alpha 1} +\frac{\overline{\dot{A}^1}}{\overline {A^1} }\left(\delta ^0_{\alpha} g_{\beta 1}-\delta ^0_{\beta}g_{\alpha 1}\right) \, ,
\end{equation}
then it not only respects  the remnant symmetry \eqref{UG-symmetry} but also is a  four- tensor under the general coordinate transformation  in the sense that 
\begin{equation}
G_{\alpha \beta} =\Lambda ^{\alpha ^\prime}_{\alpha}  \Lambda ^{\beta ^\prime}_\beta G_{\alpha ^\prime \beta ^\prime} \, .
\end{equation}
Consequently, we can construct proper scalars with the contractions of $G_{\alpha \beta}$. This is as far as we can go with
the metric perturbations and their derivatives. 

In addition, the anti-symmetric tensor $\epsilon^{\alpha \beta \mu \nu}$  can be used to construct  the dual of $G_{\alpha \beta}$ defined via
\ba
\tilde G^{\mu \nu} \equiv \epsilon^{\alpha \beta \mu \nu } G_{\alpha \beta}
\ea
which is a four-tensor too.  

In conclusion, our building blocks to construct the action in matter sector 
in unitary gauge  are $g^{00}, G_{\mu \nu}, \tilde G^{\mu \nu}$. The other building blocks like the extrinsic curvature $K_{ij}$ are geometric in nature and do not come from the matter sector.  
Since we work in decoupling limit in which the higher derivative terms are
neglected, we do not consider the contribution of geometric building blocks like $K_{ij}$ or their mixings with $g^{00}, G_{\mu \nu}, \tilde G^{\mu \nu}$. Below we justify the validity of the decoupling assumption.

The most general action for the matter sector perturbations,  up to quadratic order in perturbations, in unitary gauge are 
\begin{align}
\label{action-UG}
S=\int d^4 x \sqrt{-g}\Bigg[& \Lambda + \alpha _0 g^{00}+\frac{c_0}{4}\left(\delta g^{00}\right)^2-\frac{1}{4}M_1\delta \left( G^{\alpha\beta}G_{\alpha\beta}\right)-\frac{1}{4}M_2 \delta \left( G^{\alpha\beta}G_{\alpha\beta}\right)^2 \nonumber \\
&-\frac{1}{4}M_3 \delta \left(G^{\alpha\beta}\tilde{G}_{\alpha\beta}\right) -\frac{1}{4}M_4 \delta \left(G^{\alpha\beta}\tilde{G}_{\alpha\beta}\right)^2+\frac{1}{2}\lambda _1\delta g^{00}\delta \left( G^{\alpha\beta}G_{\alpha\beta}\right) 
\nonumber\\
&+\frac{1}{2}\lambda _2\delta g^{00} \delta \left(G^{\alpha\beta}\tilde{G}_{\alpha\beta}\right) + ... \Bigg], 
\end{align}
in which it is understood that the indices for the four-dimensional tensors are raised and lowered via $g_{\mu \nu}$
and $g^{\mu \nu}$, i.e.  $G^{\alpha\beta}=g^{\alpha \mu}g^{\beta \nu}G_{\mu \nu}$ and  similarly for $\tilde{G}_{\alpha \beta}$. 

The terms $\Lambda$ and $\alpha_0$ are fixed from the tadpole cancelation at the background level. In particular, we note that $\Lambda$ is determined by the value of the potential to support inflation while $\alpha_0 \propto \dot H$ in which $H$ is the  effective (isotropic) Hubble expansion rate.  It worth mentioning  that by putting one of $G_{\alpha\beta}$ components on the background the other terms would also make sub-dominant contributions to tadpole terms which  can be absorbed by redefinition of $\Lambda$ and $\alpha_0$. Therefore, the symbol $\delta$ behind products of $G_{\alpha\beta}$ etc means that we look at the  perturbations of the corresponding quantities, excluding their background values. 

The couplings $M_1, M_2, M_3, M_4$ and $\lambda_1, \lambda_2$ 
 are left undetermined in the spirit of EFT.  Note that in writing the action 
 we have kept terms to leading orders of derivatives, terms with higher
orders of derivatives are suppressed as long as we are working in low energy.  However, we note that the terms containing $M_2$ and $M_4$ are higher orders in derivatives respectively 
compared to $M_1$ and $M_3$ and are non-renormalizable. Therefore, in principle,  they can also be ignored to leading order of EFT analysis. However, we keep these two terms which are still leading compared to other higher derivative terms encoded in ...   which can have interesting effects for the anisotropy power spectrum. 

As usual the unitary action given above represents the action in the matter sector. In addition to this, we also have the usual gravitational action given by the Einstein-Hilbert term. However, we do not elaborate on this part as we will be
working on the decoupling limit in which the gravitational back-reactions are suppressed to leading order in slow-roll parameters as we will justify later on.

Before concluding this Section, it is instructive to compare our results with the well-studied model of anisotropic inflation 
\cite{ Watanabe:2010fh, Bartolo:2012sd} based on Maxwell theory with a time-dependent 
(actually $\phi$-dependent) gauge kinetic coupling:
\ba
\label{Maxwell}
L_{\mathrm{Maxwell}} = -\frac{f(\phi)^2}{4} F_{\mu \nu} F^{\mu \nu} \, .
\ea
As mentioned before, in order for the background electric field to contribute a nearly constant energy density to
total energy, we require $f(\phi) \propto a(t)^{-2}$,  yielding $\overline {\dot A^1} = H \overline {A^1}$.  Therefore,
\begin{align}
F_{\mu\nu}&=\partial _\mu A_\nu - \partial _\nu A_\mu , \nonumber \\
&=\overline{A^1} \left(\partial _\mu g_{\nu 1}-\partial _\nu g_{\mu 1}\right)+\dot{\overline{A^1}}  \left( \delta ^0_\mu g_{\nu 1} - \delta ^0_\nu g_{\mu 1}\right), \nonumber \\
&=\overline{A^1} G_{\mu\nu},
\end{align}
where we have used the fact that $ A_\mu = g_{\mu \nu}A^\nu = g_{\mu 1}\overline{A^1}$. 

Now comparing the Lagrangian
Eq. (\ref{Maxwell}) to our general  action Eq. (\ref{action-UG}) and using the above relation between  $F_{\mu \nu}$ and
$G_{\mu \nu}$ yields  
\begin{equation}
M_1=f^2\left(\overline {A^1}\right)^2 \propto a^{-2}, \qquad M_2=M_3=M_4=\lambda _1=\lambda _2=c_0=0.
\end{equation}
It is very interesting that the anisotropic inflation based on Maxwell theory is such a simple model compared to
general possibilities encoded in Eq. (\ref{action-UG}).


\section{The Goldstone Bosons}
\label{Goldstone sec.}

The action (\ref{action-UG}) are obtained in unitary gauge  defined such that  $\delta \phi=\delta A^\mu=0$. As usual in EFT approach, we can leave this gauge to any arbitrary coordinate system in which the full four-dimensional diffeomorphism invariance is explicit. This requires  the appearance of Goldstone bosons $\pi^\mu$
\ba
\label{pi-mu-transformation}
x^\mu \rightarrow {x^\mu}'=   x^\mu + \pi^\mu \, .
\ea
On the physical grounds, we expect to have more than one Goldstone bosons.
The Goldstone boson $\pi^0$ is associated with the breaking of time diffeomorphism which is used to set $\delta \phi=0$ in unitary gauge. The nature of $\pi^0$ is the same as in \cite{Cheung:2007st}: liberating the time coordinate, we introduce the scalar field $\pi^0(x^\mu)$ which encodes the fluctuations of the inflaton in any coordinate system. In addition, restoring the $\delta A^\mu$ fluctuations, we expect to introduce the Goldstone bosons $\pi^i$.  This suggests we will have three more Goldstone bosons. However, as we shall see, fixing the $U(1)$ gauge  will reduce this to two independent Goldstone bosons, which are the number of transverse polarization degrees of freedom of the gauge field fluctuations.

\subsection{The Quadratic Action of Goldstone Fields}

Now we can restore the Goldstone bosons and  perform the so-called Stueckelberg trick. 
 We also work in  the  decoupling limit in which the metric perturbations are neglected. 
 Also to simplify the notation, we drop the overline over $A^1$, so from now on $A^1$ simply stands for $\overline {A^1}$. 
 
 Upon restoring the Goldstone  bosons $\pi^\mu$ we have
\begin{align}
\delta g^{00}&\rightarrow 2\dot{\pi}^0+a_i^{-2}(\pi ^0_{,i})^2-(\dot{\pi}^0)^2, \\
g_{11}&\rightarrow g _{11}+2a_1^2\pi ^1_{,1}+2a_1^2 \pi ^0_{,1}\dot{\pi}^1+2a_1^2\pi ^i_{,1}\pi ^1_{,i}-(\pi ^0_{,1})^2+a_i^2 (\pi ^i_{,1})^2, \\
g_{01}&\rightarrow -\pi ^0_{,1}+a_1^2 \dot{\pi}^1+\mathcal{O}(\pi ^2), \\
g_{1i}&\rightarrow g _{1i}+a_i^2 \pi ^i_{,1}+a_1^2\pi ^1_{,i}+\mathcal{O}(\pi ^2) \, .
\end{align}
in which the notation ``$,i $ '' here and below denotes $\partial_i$, for example $\pi^j_{, i} =\partial_i \pi^j$ and so on.

Equipped with the above transformation rules of $\delta g_{\alpha \beta}$, we can calculate the quadratic action 
(\ref{action-UG}) in terms of the Goldstone fields.  The key to simplify the analysis is that we should not get any Goldstone field from indices which are contracted in Lorentz invariant manner.  For the contraction 
$G^{\alpha\beta}G_{\alpha\beta}$ we obtain 
\begin{align}
\label{G2}
G^{\alpha\beta}G_{\alpha\beta}\rightarrow & -2 a^{-2}\Big[ \partial _0\left( \overline{g}_{i\gamma ^\prime}\Lambda ^{\gamma ^\prime} _1\right) + \partial _i \Lambda ^0_1+\frac{\dot{A}^1}{\dot{A}}\left(\overline{g}_{i\gamma ^\prime} \Lambda ^{\gamma ^\prime}_1\right) \Big]^2
+ a^{-4}\left[ \partial _i \left(\overline{g}_{j\gamma ^\prime} \Lambda ^{\gamma ^\prime}_1\right)-\partial _j \left(\overline{g}_{i\gamma ^\prime} \Lambda ^{\gamma ^\prime}_1\right)\right]^2 \nonumber \\
&\rightarrow -2a^2\Big(2H+\frac{\dot{A}^1}{A^1}\Big)^2-4 \Big(2H+\frac{\dot{A}^1}{A^1}\Big) \Big( \delta \dot{X}_1+\pi ^0_{,11}+\frac{\dot{A}^1}{A^1}\delta X_1\Big) - 2a^{-2}\left(\delta \dot{X}_i\right)^2 \nonumber \\
&-2a^{-2}\left(\pi ^0_{,1i}\right)^2 -2 a^{-2}\big(\frac{\dot{A}^1}{A^1}\big)^2 \Big(\delta X_i\Big)^2 -4 a^{-2}\delta \dot{X}_i\pi ^0_{,i1} -2a^{-2} \frac{\dot{A}^1}{A^1}\frac{d}{dt}\left(\delta X_i\right)^2\nonumber \\
&-4a^{-2}\frac{\dot{A}^1}{A^1}\pi ^0_{,1i} \delta X_i +2a^{-4}\left[\Big(\delta X_{i,j}\Big)^2-\left(\delta X_{i,i}\right)^2\right],
\end{align}
in which we have defined  $\overline{g}_{\gamma i}\Lambda ^\gamma _1 \equiv \overline{g}_{1 i} + \delta X_i $
or
\ba
\delta X_i \equiv a^2 \partial_1 \pi^i \, .
\ea
The above equations indicate that it is $\delta X_i$ and not $\pi^i$ itself which is physical. This is a consequence of
our remnant symmetry Eq. (\ref{UG-symmetry}) which somewhat singles out $\partial_1$ operation in the sense that 
$\partial_i \xi_T=0$ while $\partial_\alpha  \xi_L$ is related to $\partial_1 \xi_L$. 

Similarly, for the contraction  $\tilde{G}^{\alpha\beta}G_{\alpha \beta}$ we obtain 
\begin{align}
\frac{1}{4}\tilde{G}^{\alpha\beta}G_{\alpha \beta}&=\frac{1}{4}\epsilon ^{\alpha \beta \gamma \delta} G_{\alpha\beta} G_{\gamma \delta} \nonumber \\
&\rightarrow 2\epsilon ^{1jk}\Big(2H+\frac{\dot{A}^1}{A^1}\Big) {\delta X_{k,j}} + 2\epsilon ^{ijk} {\delta X_{k,j}}\Big(\pi ^0_{,1i}+\delta \dot{X}_i+\frac{\dot{A}^1}{A^1}\delta X_i\Big),
\end{align}
where we use the convention that $\epsilon^{0123}=1$ and define $\epsilon ^{ijk}\equiv \epsilon ^{0ijk}$. 

The other terms in the action (\ref{action-UG}) can be evaluated in terms of the Goldstone bosons similarly. Combining all terms, the  full second order action written in terms of the physical fields 
$\pi^0$ and $\delta X_i$ is obtained to be 

\begin{align}
\label{S-total}
S=\int d^4x \sqrt{-g}\Bigg\lbrace & \alpha _0 \left[ -\left(\dot{\pi}^0\right)^2 +a^{-2}\left(\pi ^0_{,i}\right)^2\right] + c_0 \left( \dot{\pi}^0\right)^2 \nonumber \\
&+\frac{1}{2}M_1a^{-2}\Bigg[ \left(\delta \dot{X}_i\right)^2 + \left(\pi ^0_{,i1}\right)^2 + (\frac{\dot{A}^1}{A^1})^2 \left(\delta X_i\right)^2 +2 \delta \dot{X}_i\pi ^0_{,1i} + \frac{\dot{A}^1}{A^1}\frac{d}{dt}\left(\delta X_i\right)^2 \nonumber \\
& +2\frac{\dot{A}^1}{A^1}\pi ^0_{,1i}\delta X_i-a ^{-2}\left( (\delta X_{i,j})^2 - (\delta X_{i,i})^2 \right) \Bigg] 
\nonumber \\
&-\dot{M_1}(2H+\frac{\dot{A}^1}{A^1})\pi ^0 ( \delta \dot{X}_1+\pi ^0_{,11}+\frac{\dot{A}^1}{A^1}\delta X_1) \nonumber \\
&-4M_2(2H+\frac{\dot{A}^1}{A^1})^{{2}}\Bigg[ \left(\delta \dot{X}_1\right)^2 + \left(\pi ^0_{,11}\right)^2 + (\frac{\dot{A}^1}{A^1})^2 \left( \delta X_1\right)^2 +2\delta \dot{X}_1\pi^ 0_{,11} \nonumber \\
&+\frac{\dot{A}^1}{A^1}\frac{d}{dt}\left(\delta X_1\right)^2 + 2\frac{\dot{A}^1}{A^1}\pi ^0_{,11}\delta X_1\Bigg]  +2M_3 \epsilon ^{ijk} \delta X_{j,k}( \pi ^0_{,1i}+\delta \dot{X}_i+\frac{\dot{A}^1}{A^1}\delta X_i)\nonumber \\
&-M_4a^4 ( 2H+\frac{\dot{A}^1}{A^1})^2\epsilon ^{1jk}\epsilon ^{1lm}\delta X_{j,k}\delta X_{l,m}-4\lambda _1 \dot{\pi}^0 ( 2H+\frac{\dot{A}^1}{A^1} ) ( \delta \dot{X}_1+\pi ^0_{,11}+\frac{\dot{A}^1}{A^1}) \nonumber \\
&-8\lambda _2 (2H+\frac{\dot{A}^1}{A^1})\dot{\pi}^0 \epsilon ^{1ij}\delta X_{i,j} {-2}\dot{M}_3\epsilon ^{1jk} (2+\frac{\dot{A}^1}{A^1H})H\pi ^0 \delta X_{j,k}\Bigg\rbrace .
\end{align}
The above action have many terms. However, in the spirit of EFT we are interested in low energy (comparing to the cut off of EFT) behavior of this system. Therefore, we may adopt the Wilsonian view here and only take into account terms with least number of derivatives so we consistently discard terms with 3 or higher number of derivatives of $\pi^0$ and $\delta X_i$
in the following analysis.

\subsection{The Free Fields}

Here we read off the action of the free fields from the total action (\ref{S-total}) and their wave functions.  

First we start with $\pi^0$ field which is simpler. 
After some integrations by parts the action for $\pi ^0$ field is obtained to be 
\begin{align}
\label{pi0-action}
S^{\pi ^0}_2=\int d^4x\sqrt{-g}\Bigg\lbrace& (-\alpha _0) \left[ \left(\dot{\pi}^0\right)^2\left(1-\frac{c_0}{\alpha _0}\right) - a^{-2}\left(\pi ^0_{,i}\right)^2\right]\nonumber \\ &+{(2+n)H}\left[\dot{M}_1+2\left(\dot{\lambda}_1+3H\lambda _1\right)\right] \left(\pi ^0_{,1}\right)^2 +\ldots \Bigg\rbrace , 
\end{align}
in which $\ldots$ denotes  terms with  higher number of derivatives.  Note that $\alpha_0 \propto \dot H  <0$ so the kinetic energy has the proper sign. 

The free wave function of $\pi ^0$ with the Minkowski  initial conditions deep inside the horizon  is
\begin{equation}
\label{pi0-wave}
\pi ^0(k)=\frac{H}{2k^{3/2}}\sqrt{\frac{\pi}{2c_s|\alpha _0|}}(-kc_s\tau)^{3/2}H^{(1)}_{3/2}(-kc_s\tau),
\end{equation}
where we defined the sound speed of $\pi^0$ fluctuations 
\begin{equation}
\label{cs-eq}
{c_s^{-2}=1-\frac{c_0}{\alpha _0}}.
\end{equation}
As expected from the discussions of \cite{Cheung:2007st} the coefficient $c_0$ controls the sound speed of the $\pi^0$ fluctuations. This can arise for example in the models of k-inflation \cite{ArmendarizPicon:1999rj, Garriga:1999vw}
or DBI inflation  \cite{Alishahiha:2004eh} as is well-understood in inflation literature. Now
the interesting effect is that we can extend the DBI-type model to anisotropic inflation with gauge fields. This may have motivations from string theory in which the world volume of a mobile D3 brane contains the $U(1)$ gauge fields.   For a model of anisotropic inflation with DBI type action  see \cite{Dimopoulos:2011pe}.

Note that we have discarded the contributions from the terms in second line of Eq. (\ref{pi0-action})  in the free wave function. 
In principle we can include the contributions of these terms in curvature perturbations power spectrum via their corrections to $\pi^0$ free wave function. However, their contribution is sub-leading as follows.  If we look at the terms containing  $M_1$ and $\lambda_1$
we see that these terms come from perturbing the term $G_{\alpha\beta}G^{\alpha\beta}$. Therefore, we will have a contribution from these terms to the energy content during inflation. However, as the dominant source of the background expansion  comes from the inflaton sector,  the contribution of these terms to $\Lambda$ and the coefficients of tadpole terms should be small. This means that $M_1H^2<<|\alpha _0|$ so we can neglect the contributions of the terms in second line of  Eq. (\ref{pi0-action}) to leading order.

Now we calculate the free wave functions of the $\delta X_i$ fields. 
As usual,  we want our canonical fields to be massless. The action of the free $\delta X_i$ fields  has the following general form 
\begin{equation}
S=\frac{1}{2}\int d^4x \sqrt{-g}\left(\mathcal{N}\dot{\phi}^2+m^2\phi ^2+\ldots \right), 
\end{equation}
in which ${\cal N}$ is a time-dependent normalization and $\phi$ collectively represents $\delta X_i$ fields. 
Correspondingly, the canonically normalized field is given by  $\phi _c=\sqrt{\mathcal{N}}\phi$ and the condition for $\phi_c$ to be massless is 
\begin{equation}
-\frac{1}{2}\dot{\mathcal{N}}^2+3H\mathcal{N}\dot{\mathcal{N}}+\mathcal{N}\ddot{\mathcal{N}} +2\mathcal{N}m^2=0. \label{mass-condition}
\end{equation}

Looking at the actions for $\delta X_i$, we see that $\delta X_2$ and $\delta X_3$ have the same coefficients  which are
different than those of $\delta X_1$. Let us first  consider   $\delta X_2$ and $\delta X_3$ which are easier. We find that
the coefficients ${\cal N}$ and $m^2$ for $\delta X_2$ and $\delta X_3$ fields are proportional to the unknown coupling of EFT $M_1$. So far our analysis was generic with no assumptions on the time scaling of EFT coefficients. However,  experience from the previous specific models of anisotropic inflation and the structure of our in-in integrals suggest that it is very reasonable to
assume a time scaling like 
\ba
\label{M1-scaling}
M_1= \overline M_1 a^{s_1} \, ,
\ea
with $\overline M_1$ and $s_1$  being constant. This is not the most general functional form of $M_1$ but it is generic enough for our purpose which captures all the models of anisotropic inflation
studied so far. In addition, as we shall see, this scaling with time is actually what our in-in integrals suggest for interesting physical results.  With similar reasoning, we also assume 
\begin{equation}
\frac{\dot{A}^1}{A^1}=nH \, ,
\end{equation} 
with $n$ being a constant.  

With these scaling ansatz for $M_1$ and $A^1$  we obtain
\begin{equation}
\mathcal{N}=M_1a^{-2}=\overline{M_1}a^{-2+s_1}, \qquad m^2=\overline{M}_1(n^2-n-ns_1)a^{-2+s_1} {H^2} \, .
\end{equation}
Plugging these into Eq. \eqref{mass-condition} we obtain,
\begin{equation}
s_1^2+s_1(2-4n)+4n^2-4n-8=0 \Rightarrow s_1=-1+2n \pm 3. 
\label{condition-m1}
\end{equation}
The above equation gives a relation between $n$ and $s_1$, but does not fix them individually. We can fix $s_1$ by 
checking the contribution of term containing $ M_1 G_{\alpha \beta} G^{\alpha \beta}$ into the background inflationary expansion. 
At the background level we have $G_{0i} = a^2  ( 2 H + \frac{\dot{A}^1}{A^1})  \delta_{i 1} $  with the other components being zero. As a result, at the background level we have
\ba
\label{M1-back}
-\frac{1}{4} M_1 \overline G_{\alpha \beta} \overline G^{\alpha \beta} = \frac{1}{2} M_1 a^2   \big( 2 H + \frac{\dot{A}^1}{A^1} \big)^2 \, .
\ea
The above term contributes to the background inflation expansion  via renormalizing the cosmological constant term. In order to have a long enough period of inflation with 
small amount of anisotropy, we require that the above term to be nearly time-independent so it only modifies the effective 
cosmological constant. This requires that $M_1 \propto a^{-2}$. Comparing to ansatz provided in Eq. (\ref{M1-scaling}) this yields $s_1 =-2$. Consequently, from Eq. (\ref{condition-m1}) we obtain $n=1$ and $n=-2$. The latter corresponds 
to $A_1 = \mathrm{constant}$,  yielding a zero electric field energy density. Therefore, it is a trivial solution and we conclude that the only allowed value  is $n=1$.  Having said that, in order to keep track of the role of parameter $n$ we
leave it undetermined, but we will impose the conclusion  $n=1$ in our final results. 

Now we look at the free action for $\delta X_1$ field. In addition to common terms similar to the free actions of  $\delta X_2$
and $\delta X_3$, we have a new contribution from $M_2$. Motivated from the above discussions, we assume  $M_2=\overline{M}_2a^{s_2}$ with $\overline M_2$ and $s_2$ being constants. 
In addition, we assume that the  time scaling of $M_1a^{-2}$ is equal to $M_2$ since both of them contribute to the kinetic energy 
of $\delta X_1$ field and  we do not want one of them to dominate  over the other  during inflation. We will show momentarily that this is indeed a consistent assumption. In conclusion, for $\delta X_1$ field we have,
\begin{equation}
\mathcal{N}=M_1a^{-2}-8M_2H^2(2+n)^2, \qquad s_2=s_1-2
\end{equation}
and with $s_2=s_1-2$ the mass term $m$ is given by
\begin{equation}
m^2 = \mathcal{N}H^2 (n^2-n-ns_1).
\end{equation}
Plugging these in Eq.  \eqref{mass-condition} we obtain,
\begin{equation}
-\frac{1}{2}(s_1-2)^2+3(s_1-2)+(s_1-2)^2+2(n^2-n-ns_1)=0.
\end{equation}
Note that this is exactly the same equation as \eqref{condition-m1} which shows that our assumption on taking 
$M_2 \propto M_1a^{-2}$ was consistent. In conclusion, for the scaling of $M_2$ we have $s_2 =-4$. 

As mentioned before, it seems we will get three independent Goldstone bosons from $\delta X_1, \delta X_2$ and $\delta X_3$ fields. However, we should recall that these fields are associated with restoring the $\delta A^\mu$ field after liberating ourselves from the unitary gauge. Therefore, we should be careful of the remnant $U(1)$ gauge symmetry to be imposed on $\delta A^\mu$ fluctuations in {\it any} coordinate system. 
To see this more specifically, suppose we move from the unitary gauge to the arbitrary coordinate after restoring the Goldstone bosons $\pi^i$  as given in Eq. (\ref{pi-mu-transformation}). Then 
the gauge field perturbations transform as
\ba
\delta A^i \rightarrow  {\delta A^i}'  = {\delta A^i} + \partial_{1} \pi^i  A^{1} =   {\delta A^i} + \delta X_i  A^{1} \, .
\ea
Already in writing the action in unitary gauge we assumed that the $U(1)$ gauge is fixed. Here after restoring the coordinate invariance we should check the presumed $U(1)$ gauge condition.  

Now to fix the $U(1)$ gauge  we impose the Coulomb-radiation gauge
in which $A^0 = \partial_i A^i=0$. Combining with the above coordinate transformation, this requires  
\ba
\label{cond1}
\partial_i \delta X_i=0 \, .
\ea
Now decompose $\delta X_i$ into its longitudinal and transverse parts as follows
\ba
\label{LT-dec}
\delta X_i = \partial_i \delta  X_L + \delta X_{ T i}    \quad , \quad \partial_i  \delta X_{T  i} =0 \, .
\ea
Combining Eq. (\ref{cond1}) with Eq. (\ref{LT-dec}) we conclude that $\nabla^2 \delta X_L=0$. With the appropriate boundary conditions at infinity, this yield $\delta X_L=0$. Therefore, we come to the important conclusion that the longitudinal part of
$\delta X_i$ perturbations are not physical and only the transverse parts of $\delta X_i$ are physical. These two physical degrees of freedom are indeed the two transverse polarization of the $U(1)$ gauge field as anticipated.

Our job now is to find the action of the free fields $\delta X_T$. Using the relation 
\begin{equation}
\epsilon ^{ijk}\delta X_{j,k}\delta \dot{X}_i=\frac{1}{2}\left[ \partial _0 \left(\epsilon ^{ijk} \delta X_I \delta X_{j,k}\right) - \partial _k \left( \epsilon ^{ijk} \delta X_i \delta \dot{X}_j\right)\right],
\end{equation}
we obtain the following  action for free  $\delta X_{T i}$ fields 
\begin{align}
S_2^{X_T}=&\int d^4x \sqrt{-g}\Big \lbrace \frac{1}{2} \left(\delta \dot{X}^c_{ T i}\right)^2 -\frac{1}{2} a^{-2} \left(  \delta X^c_{Ti, j}\right)^2 - a^{-2}\frac{4\overline{M}_2H^2(2+n)^2}{\overline{M}_1-8\overline{M}_2H^2(2+n)^2}\left(\delta X^c_{T1,i}\right)^2 \nonumber \\
&+{2}M_3H (n{-}\frac{3}{2}{-}\frac{\dot{M}_3}{2HM_3}) \epsilon ^{ijk}\delta X_{Ti}\delta X_{Tj,k} - M_4 a^4 H^2 (2+n)^2 \epsilon ^{1ij}\epsilon ^{1kl} \delta X_{Ti,j}\delta X_{Tk,l}\Big\rbrace , 
\label{action-X}
\end{align}
in which the canonically normalized fields $\delta X^c_{T i}$ are defined via
\begin{align}
\delta X^c_{ T j}&=\sqrt{M_1} a^{-1}\delta X_{T j}, \qquad j=2,3 \\
\label{normalization}
\delta X^c_{ T  1}&=\sqrt{M_1a^{-2}-8M_2H^2(2+n)^2}\delta X_{T1}.
\end{align}
Furthermore, it is convenient to decompose $X^c_{ T  i}$ in terms of the gauge field polarization base $\epsilon ^s_i(k)$ in Fourier space 
\begin{equation}
\delta X^c_{Ti}=\sum _s \delta X^{c\,  (s)}_{T }(k,t)\epsilon ^s_i(k)
\end{equation} 
where $\epsilon ^s$ denotes the polarization vector and satisfies certain orthogonality relations. We can use either the linear polarization 
base with $s=1,2$ or the circular (helicity) base with $s =\pm$ but at this stage we do not fix the base. 

Imposing the Minkowski  initial conditions deep inside the horizon we obtain 
\begin{equation}
\label{Xc-wave}
\delta X^{c(s)}_{T } =-\frac{H\sqrt{\pi}}{2k^{3/2}}(-k\tau )^{3/2} H^{(1)}_{3/2}(-k\tau) \, ,
\end{equation}
and finally,
\begin{align}
\label{XT1-wave}
\delta X_{T1}&=\frac{a^2}{\sqrt{\overline{M}_1-8\overline{M}_2H^2(2+n)^2}}\sum _s \delta X^{c\,  (s)}_{T } \epsilon ^s _1 (k) \\
\delta X_{T j}&=\frac{a^2}{\sqrt{\overline{M}}_1}\sum _s \delta X^{c\,  (s)}_{T } \epsilon ^s _2 (k),  \quad j=2, 3 \, .
\end{align}
In obtaining the above equations we neglect $M_3$ and $M_4$ terms which modify different polarization components of gauge field. However, we  will take into account their contribution as perturbations to $\delta X_{T i}$  wave function which can also affect the anisotropic power spectrum.

\subsection{The Interactions}

Having calculated the wave functions of the free fields, here we obtain the interaction between the fields which are in the form of exchange vertices. After integration by parts and noting that $\delta X_{i,i}=0$, we obtain,
\begin{align}
\label{S-int}
S_2^{int}=\int  d^4x \sqrt{-g}\Bigg\lbrace &-\dot{M}_1(2+n)H \pi ^0 \delta \dot{X}_{T1}-8M_2(2+n)^2H^2 \pi ^0_{,11}\left( \delta \dot{X}_{T1} + nH \delta X_{T1} \right) \nonumber \\
&+2M_3\epsilon ^{ijk}\pi ^0_{,1i}\delta X_{Tj,k}-4\lambda _1 H (2+n)\dot{\pi}^0 \left(\delta \dot{X}_{T1}+nH \delta X_{T1}\right) \nonumber \\
&-8\lambda _2 (2+n)H \epsilon ^{1jk}\dot{\pi}^0 \delta X_{Tj,k}{-2}\dot{M}_3 (2+n)H \epsilon ^{1jk} \pi ^0 \delta X_{Tj,k}
 \nonumber \\
&{-\dot{M}_1(2+n)nH^2 \pi ^0 \delta X_{T1}}\Bigg\rbrace .
\end{align}
Fortunately many  terms in the interaction Lagrangian above are irrelevant for low energy EFT studies. 
The terms involving $M_2, M_3, \dot M_3$ and $\lambda_2$ are suppressed on super-horizon scales due to presence of spatial partial derivatives so they can be discarded in low energy EFT limit. However, one may argue that we do not know the scaling of coefficients $M_3$ and $\lambda_2$ so if their time-dependence is singular, i.e. containing positive power of $a(t)$, then their
contributions may not be so obviously suppressed compared to terms containing $M_1$ and $\lambda_1$. To answer this concern 
we estimate the time scaling of these coefficients. First we note that $M_2 \sim M_1 a^{-2}$ so the term in Eq. (\ref{S-int})
containing  $M_2$ are highly suppressed compared to terms containing $M_1$ so it can safely be ignored. 
To obtain the scaling of $\lambda_1$ we note that the term containing $\lambda_1$  comes from perturbing $g^{00}G^2$ which yields 
\begin{equation}
\label{lambda1-scale}
\lambda_1 g^{00}GG=- \lambda_1 \overline{G^2}- \lambda_1\delta \left(G^2\right)- \lambda_1 \delta g^{00}\delta \left( G^2\right), 
\end{equation}
in which $\overline{G^2}$ represents the background value of ${G^2}$. 
Hence the above term  gives corrections to cosmological constant and also to the coefficient of $\delta ( G_{\alpha\beta}G^{\alpha\beta}) $ which is $M_1$. Now noting that $\overline{G^2} \propto a^{2}$, and  in order for the effective cosmological constant to stay nearly constant,  we require that $\lambda_1 \propto  M_1 \propto a^{-2} $.  Therefore, the interaction in Eq. (\ref{S-int}) containing $\lambda_1$ is as relevant as those of $M_1$. 
As for $\lambda_2$ we see that the  term containing $\lambda_2$  originates from perturbing $g^{00}G_{\alpha\beta} \tilde{G}^{\alpha\beta}$,
\begin{equation}
\lambda_2 g^{00}G\tilde{G}= \lambda_2 \delta g^{00} \overline{G\tilde{G}}- \lambda_2 \delta \left(G\tilde{G}\right).
\end{equation} 
Hence, the last term above also contributes  to $M_3$ so the scaling of $\lambda_2$ with time must be the same as $M_3$.
However, as we will argue in next Section, $M_3$ scales like $a^{-5}$ so the interactions containing  $\lambda_2$ and $M_3$
are highly suppressed.

After some integration by parts, and going to conformal time $\tau$, the interaction Lagrangian becomes  
\begin{equation}
S^{\mathrm{int}}_2=\int  d \tau d^3x \left(L_1+L_2+L_3\right),
\end{equation} 
in which,
\begin{align}
\label{L1}
L_1&=a^2\Big[ 2\overline{M}_1 (n+2)(n-1)H^3\Big]  \pi ^0 \delta X_{T1}, \\
\label{L2}
L_2&=-a\Big[4\overline{\lambda}_1 H^2 {n}(2+n)+2H^2(2+n)\overline{M}_1\Big] {\pi ^0}^\prime \delta X_{T1}, \\
\label{L3}
L_3&=-4\overline{\lambda}_1H(2+n) {\pi ^0}^\prime \delta X_{T1}^\prime .
\end{align}
where a $^\prime$ denotes derivative with respect to conformal time and we have defined the scaling of $\lambda_1$ as
$\lambda_1 =  \overline{\lambda}_1 a^{-2} $ as discussed above.  

Before concluding this section, here we discuss the  validity of decoupling limit which was used to simplify the analysis
significantly and also estimate the UV cutofo of the theory due to strong interactions of $\delta X_i$. Let us first start  with
the justification of our decoupling assumption.  The leading term with least number of derivatives which mixes $\pi ^i$ field with the metric comes from,
\begin{equation}
M_1  H^2 a^{-2}\delta g_{i1}\delta X_i \, .
\end{equation}
Related to  the canonically normalized fields  $\delta X^c_i\sim\sqrt{M_1}a^{-1}\delta X_i$ and 
$\delta g^{c}_{1 i} \sim \delta g_{1 i} M_P^{-1}$,  this interaction becomes 
\begin{equation}
\overline M_1^{\frac{1}{2}} M_P^{-1} H^2 \delta g^{c}_{i1}\delta X^c_i.
\end{equation}
Comparing this term with kinetic term $(\dot {\delta {X^{c}_i}})^2$ we are able to estimate the  mixing energy,
\begin{equation}
E_{\mathrm{mix}}\sim \frac{\overline{M}_1^{1/4} H}{M_P}.
\end{equation}
As we will see in next Section, $\overline{M_1}$ controls the fraction of energy density of gauge field (see 
Eq. \ref{M1-relation}) to the total energy density which is very small.  Therefore this mixing energy lies well outside horizon and for energies greater than $E_{\mathrm{mix}}$, we can safely neglect mixing of $\delta X_i$ with gravity.

Now we may estimate the UV cutoff of the theory due to strong interactions of $\delta X_i$. The cutoff of theory due to strong interactions of $\pi ^0$ is estimated in \cite{Cheung:2007st}. Obviously, the lower cutoff will be the cutoff of our theory at which our effective field theory fails to be weakly interacting.
For simplicity, we drop numerical factors and scale factors $a(t)$ in following discussions. From \eqref{G2} and \eqref{action-UG} it is clear that the first non-trivial interaction between $\delta X_i$ arises from $M_2\left(G_{\alpha\beta}G^{\alpha \beta}\right)^2$:
\begin{equation}
\overline{M}_2H\delta\dot{X}_1(\partial _i \delta X)^2.
\end{equation}
As we shall see in next Section, this operator will generate a non-trivial sound speed  $c_v$ for gauge field fluctuations, see Eq. (\ref{cv-eq}).  One way to deal with this non-trivial sound speed is to re-scale $x^i\rightarrow \tilde{x}^i={x^i}/{c_v}$. With this rescaling we may define a $\tilde{\partial} _\mu=(\partial _0, c_v\partial _i)$ and make our free theory to be explicitly  Lorentz-invariant.  Note that the Lagrangian changes as $\mathcal{L}\rightarrow c_v^3 \mathcal{L}$.

{Now our operator is a dimension six operator and upon canonical normalization and rescaling sound speed it becomes
\begin{equation}
\frac{\overline{M}_2}{\overline{M}_1^{3/2}}Hc_v^{-5}\delta \dot{ X}^c(\tilde{\partial} _i\delta X^c)^2. 
\end{equation}
Hence the cutoff of the theory becomes,
\begin{equation}
E^2_c\sim \frac{\overline{M}_1^{3/2}c_v^5}{\overline{M}_2H}.
\end{equation} }
 
Note that if $\overline{M}_2\rightarrow 0$ then $E_c\rightarrow \infty$. This is consistent with our intuition since dropping $M_2\left(G_{\alpha\beta}G^{\alpha \beta}\right)^2$ there would be no self-interaction for $\delta X$ coming from the term $M_1\left(G_{\alpha\beta}G^{\alpha \beta}\right)$ and hence the UV cutoff should be as high as $M_P$.  However, note that there are self-interactions in other terms of Lagrangian, for example in terms proportional to $M_4$, and dropping $M_2$ term, the UV cutoff of the theory should be determined with this operator. Here we neglected 
the $M_4$ operator since as it will become clear in next Section, unlike $M_2$, it is not relevant for producing observable signatures. As the final comment, as one might expect, lowering the speed of sound tends to make our theory more strongly interacting lowering the value of   $E_c$.

\section{The Anisotropic Power Spectrum}
\label{Anisotropy sec.}

Having obtained the wave functions of the free theory and the interaction Lagrangians we are able to calculate 
the anisotropy corrections to the curvature perturbations power spectrum. First, we relate $\pi^0$ to
comoving curvature perturbations $\calR$ to leading order via
\ba
\label{pi-R}
\calR = - H \pi^0 + O \left(( {\pi^0}) ^2 \right) \, .
\ea
Then to calculate the corrections to curvature perturbation power spectrum, we use the standard in-in formalism \cite{Weinberg:2005vy, Chen:2010xka, Wang:2013eqj} in which 
\begin{equation}
\label{Pij}
\delta P_{ji}=-\int ^{\tau _e}_{-\infty} d\tau _1 \int ^{\tau _1}_{-\infty} d\tau _2 \Big\langle \, \Big[ \, L_i({\tau _2}), \Big[ L_j({\tau _1}),{\pi ^{0*}(\tau _e)\pi ^{0*}(\tau _e)}\Big ] \, \Big] \, \Big\rangle ,
\end{equation}
where $\tau _e$ denotes the time of end of inflation and  $L_i$ and $L_j$ stands for either of $L_1, L_2$ and $L_3$ given in Eqs. (\ref{L1}),  (\ref{L2}) and (\ref{L3}). Note that the relation between $\calR$ and $\pi^0$ given in Eq. (\ref{pi-R}) has corrections from the direct
contributions of gauge field energy density into $\calR$. However these corrections are suppressed as we look into
leading order curvature perturbation anisotropy. 

Below we calculate the anisotropy corrections to power spectrum using Eq. (\ref{Pij}). Before doing that we mention again that the couplings $M_1$ and $\lambda_1$ play differently than the couplings $M_2, M_3$ and $ M_4$. The couplings 
$M_1$ and $\lambda_1$ appear directly in interaction Lagrangians $L_i$ so they plays the role of exchange vertices. 
The couplings $M_2, M_3 , M_4$ do not appear in $L_i$ directly, but they modify the free wave functions
of $\delta X_{T 1}$ appearing in $L_i$ so they also affect the anisotropic power spectrum.  Finally, $\lambda_2$ neither appear in $L_i$ nor modify $\delta X_{T 1}$ to leading order so it does not contribute to anisotropic power spectrum.

In order to get better insights 
about various contributions, it is helpful to look at different limits of parameter space when some couplings  are turned off and vice versa. 

\subsection{The case  $M_2=M_3=M_4=0$}

Here we consider the case where  $M_2=M_3=M_4=0$ while $M_1 $ and $\lambda_1$ are turned on. 
Also we allow for $c_0 \neq 0$. As we have seen from Eq. (\ref{cs-eq}), a non-zero $c_0$ will introduce a
non-trivial value of $c_s$ for the sound speed of $\pi^0$ fluctuations. This can arise in models such as
k-inflation \cite{ArmendarizPicon:1999rj, Garriga:1999vw} or DBI inflation \cite{Alishahiha:2004eh}.  
The coupling $M_1$ controls the kinetic energy of $\pi_T$ fluctuations. In simple models of anisotropic  inflation based 
on Maxwell theory with Lagrangian given in Eq. (\ref{Maxwell}), one has $M_1= f^2{ (A^1)}^2 = f^2 {( \dot A^1)}^2$.  On the other hand, the coupling $\lambda_1$  arises if the gauge kinetic
coupling depends on $ {\cal X} \equiv \frac{-1}{2} \partial_\mu \phi \partial^\mu \phi$ , such as in theory with
\ba
\label{fX}
L_{\mathrm{int}} = f(  {\cal X} ) F_{\mu \nu} F^{\mu \nu} \, .
\ea
One can easily check that upon perturbing $ {\cal X}$ we obtain $\delta  {\cal X} \rightarrow \delta g^{00} \dot \phi^2$ 
so $\lambda_1 \propto \dot \phi^2 a^{-2}$ in which the factor $a^{-2}$ is required to obtain the proper time
scaling of  $\lambda_1$ as discussed around Eq. (\ref{lambda1-scale}). As far as we are aware, there is no
model of anisotropic inflation in literature which has studied the effects of the coupling $\lambda_1$. This is a manifestation of the power of EFT which allows one to study different types of interactions without relying on particular models in which  different possibilities, such as the  coupling $\lambda_1$, appear naturally based on symmetry considerations.  Having said that, we would like to study in more details  the effect of the coupling $\lambda_1$ 
for anisotropic  power spectrum and bispectrum in models  such as Eq. (\ref{fX}) elsewhere.

The structure of in-in integrals is as given in Eq. (\ref{Pij}) in which there are nine possible terms to be calculated 
in the the form of $\delta P_{ij}$.  Here as an example  we illustrate how $\delta P_{11}$ is obtained. Using the form of wave functions $\pi^0$ and $\delta X_{T 1}$ given in Eqs.   (\ref{pi0-wave}),   (\ref{Xc-wave}) and (\ref{XT1-wave}) 
we obtain
\ba
\delta P_{11}&=& \left[2\overline{M}_1 (n+2)(n-1)H^3\right] ^2  \int ^{\tau _e}_{-\infty}  \frac{d\tau _1}{\tau _1 ^2 H^2} \mathrm{Im}\left[ \pi ^0(\tau _1) \pi ^{0*}(\tau _e)\right]
\nonumber \\ 
 ~~~~~~~~~~~~~ & \times &  \int ^{\tau _1}_{-\infty}  \frac{d\tau _2}{\tau _2^2H^2} \mathrm{Im}\left[ \pi ^0(\tau _2) \pi ^{0*}(\tau _e) \delta X_{T1}(\tau _2)\delta X^*_{T1}(\tau _1)\right] \nonumber \\
&=&\frac{2}{9 \alpha _0^2 k^3}  {(2+n)^2(n-1)^2 c_s(c_s+1)(c_s^2-c_s+1)\overline{M}_1H^4 N^2} \sum _s |\epsilon ^s_1(k)|^2,
\ea
where $N=-\ln (-k\tau)$ is number of e-folds when the mode $k$ leaves the horizon till the
end of inflation. 

Similarly, for other contributions we obtain 
\ba
\delta P_{12}&=&\frac{2}{3\alpha _0^2 k^3} {c_s^4 H^4(n-1)(2+n)^2\left(2n\overline{\lambda}_1+\overline{M}_1\right)N^2} \sum _s |\epsilon ^s_1(k)|^2 \\
\delta P_{21}&=&\frac{2}{3 \alpha _0^2 k^3}{c_s(c_s+1)(c_s^2-c_s+1)(n-1)(2+n)^2 \left( 2n\overline{\lambda}_1+\overline{M}_1\right)N^2}\sum _s |\epsilon ^s_1(k)|^2 \\
\delta P_{22}&=&\frac{2}{  \overline{M}_1\alpha _0^2 k^3 } c_s^4H^4(2+n)^2\left( 2n\overline{\lambda}_1+\overline{M}_1\right){^2}N^2 \sum _s |\epsilon ^s_1(k)|^2 \\
\delta P_{33}&=&\frac{32}{\overline{M_1}\alpha _0^2 k^3}{(2+n)^2c_s^4\overline{\lambda}_1^2H^4N^2} \sum _s |\epsilon ^s_1(k)|^2 
\ea
\ba
\delta P_{13}&=&\frac{8}{3\alpha _0^2 k^3} {c_s^4 H^4(2+n)^2 (n-1)\overline{\lambda}_1 N^2}\sum _s |\epsilon ^s_1(k)|^2 \\
\delta P_{31}&=&\frac{8}{3 \alpha _0^2 k^3} {c_s(c_s+1)(c_s^2-c_s+1)(2+n)^2(n-1) H^4\overline{\lambda}_1N^2} \sum _s |\epsilon ^s_1(k)|^2\\
\delta P_{32}&=&\frac{8}{ \overline{M}_1\alpha _0^2 k^3}{ c_s^4(2+n)^2H^4\overline{\lambda}_1\left(2n\overline{\lambda}_1 +\overline{M}_1\right) }  \sum _s |\epsilon ^s_1(k)|^2\\
\delta P_{23}&=&\delta P_{32}.
\ea
Adding all terms together, yields our final result for the anisotropy correction in power spectrum
\begin{align}
\label{delta P-case1}
\delta P=\frac{2 H^4c_s(2+n)^3N^2}{9 k^3\overline{M}_1\alpha _0^2} {\left(\overline{M}_1+6\overline{\lambda}_1\right) \left[ c_s^3(n+2) (\overline{M}_1 +6\overline{\lambda}_1) +(n-1)\overline{M}_1 \right]}  \sum _s |\epsilon ^s_1(k)|^2.
\end{align}
In the above expression, we have left the parameter $n$ undetermined, but as we argued below Eq. (\ref{M1-back}), 
the only allowed value is  $n=1$
which simplifies the above results to some extent.

To simplify the result further, we use the symmetry in the $yz$ plane to choose the wave number as 
\begin{equation}
\mathbf{k}=k\left(\cos \theta ,\sin \theta ,0\right) \, ,
\end{equation} 
where $\theta $ represents the angle between the wave number and the preferred direction $\hat {\bf n}$,  i.e. $\cos \theta = \widehat {\bf  k} \cdot \widehat {\bf n}$ in which in  our case  $\widehat {\bf n}$ is along the $x$ direction.  
As for the polarization vectors, we can use either the linear base or the helicity base. For the former, a convenient choice is
\ba
\label{linear-base}
\epsilon^{(1)} = (-\sin \theta, \cos \theta, 0) \quad ,\quad \epsilon^{(2)} = (0, 0, 1)  \, .
\ea
Consequently, the helicity base can be expressed in terms of the linear base as follows 
\ba
\label{circular-base}
\epsilon^{(+)} = \frac{i}{\sqrt2} ( \epsilon^{(1)} + i \epsilon^{(2)}) \quad , \quad
\epsilon^{(-)} = \frac{-i}{\sqrt2} ( \epsilon^{(1)} - i \epsilon^{(2)}) \, .
\ea
Using either base we obtain $ \sum _s |\epsilon ^s_1(k)|^2 =\sin^2 \theta $. 

Usually, we are interested in fractional change in power spectrum, $\frac{\delta P}{P_{\pi^0}}$, in which 
$P_{\pi^0}$ represents the power spectrum of the $\pi^0$ field which is 
\begin{equation}
P_{\pi ^0}=\frac{H^2}{4|\alpha _0| k^3c_s},
\end{equation}
in which $\alpha _0=-\epsilon M_P^2 H^2$ from tadpoles cancellation. Using $\delta P$ obtained in 
Eq. (\ref{delta P-case1}) we obtain
 \ba
 \label{deltaP-ratio1}
\frac{\delta P}{\delta P_{\pi^0}}= \frac{8 \overline M_1 c_s^2}{9 \epsilon M_P^2 }
 {(n+2)^3  (1+ \frac{6 \overline \lambda_1}{\overline M_1})  }
\Big[  c_s^3 (n+2) (1+ \frac{6 \overline \lambda_1}{\overline M_1}) +  (n- 1)  \Big] N^2 \sin^2 \theta  \, .
\ea 
Comparing the above expression with the amplitude of quadrupole anisotropy defined in Eq. (\ref{g*}) and taking
$n=1$ yields
\ba
\label{g*result1}
g_* = 72  \frac{  \overline M_1 c_s^5}{ \epsilon M_P^2 } \left(1+ \frac{6 \overline \lambda_1}{\overline M_1} \right )^2 N^2 \, .
\ea
As mentioned  before, the observational constraints require that $| g_*| \lesssim 10^{-2}$. This can be used to fix a combination of the parameters such as $c_s, \overline M_1, \bar \lambda_1$.  As in simple models of anisotropic inflation, we see again the $N^2$ structure of the anisotropic power spectrum. As discussed in \cite{Bartolo:2012sd}, this is a consequence of the accumulative contributions of  IR modes which have left the horizon and become classical, modifying the background anisotropy. 

Now let us apply the result above to the simple model of anisotropic inflation based in Maxwell theory 
given in Eq. (\ref{Maxwell}) with   $c_s=1$,  $\lambda _1=0$ and  with potential $V(\phi) = \frac{m^2}{2}{\phi^2}$. As mentioned before, in order for the gauge field furnish a sub-dominant but nearly constant portion of the total
energy density, the functional form of $f(\phi)$ have to be fine-tuned. As shown in 
\cite {Watanabe:2009ct} if one choses 
\ba
\label{f-form0}
f(\phi) = \exp {\left( \frac{c\, \phi^2}{2 M_P^2}  \right)} \, ,
\ea
with $c>1$ being a constant, then the system reaches the attractor solution in which 
the electric field energy density is a  sub-dominant but constant contribution to the total energy density.
Denoting the fraction of electric field energy density to total energy density by parameter $R$, we obtain
\ba
R \equiv \frac{\dot A_1^2 f(\phi)^2 a^{-2}}{2 V} \simeq \frac{I}{2} \epsilon
\ea
in which $I \equiv \frac{c-1}{c}$ and $\epsilon$ is the usual slow-roll parameter 
$\epsilon = -\frac{\dot H}{H^2}$.  Correspondingly, we can relate our $\overline M_1$ to $R$ via
\ba
\label{M1-relation}
\overline{M}_1=a^2f^2\left(A^1\right)^2=\frac{1}{9H^2}f^2a^{-2}\left(\dot{A}^1\right)^2 = \frac{2}{3}R M_P^2=\frac{1}{3}\epsilon I M_P^2.
\ea
Now plugging these values in our expressions for $g_*$ in Eq. (\ref{g*result1}) yields 
 \ba
 \label{deltaP-ratio2}
g_*= 24 I N^2  \,, \quad \quad   ( \mathrm{ Maxwell ~ theory} )
\ea
which is in exact agreements with the results obtained in  \cite{ Watanabe:2010fh,   Emami:2013bk, Abolhasani:2013zya, Abolhasani:2013bpa,  Chen:2014eua, Bartolo:2012sd, Shiraishi:2013vja}. 

From our analysis we conclude that $g_* \propto N^2$.  
Having this said, the relation $g_* \propto N^2$ was revisited in \cite{Naruko:2014bxa} 
in which the assumption of the attractor regime as employed in \cite {Watanabe:2009ct}
was dropped. This corresponds to an intermediate stage in which the system has not reached the attractor regime or the
total number of e-folds are limited so the IR modes which have left the horizon did not accumulate enough to modify the
background.
Compared to our analysis, this corresponds to imposing different time-scaling for $M_i$ and $\lambda_i$ than obtained in previous Section. For example, as we have seen before, the condition $M_1 \propto a^{-2}$ was achieved demanding that the anisotropic solution follows the isotropic background so the gauge field's contribution to total energy density 
is sub-leading but nearly constant, i.e. $R \sim I \epsilon$ as seen above. If we drop this assumption, then $M_1$ and other couplings  will acquire a different time-dependence than we used above, yielding  a more complicated $N$-dependence in $g_*$. As we mentioned before, we are interested in physically well-motivated situation in which the system has reached the attractor regime and our assumptions on the time scaling of various couplings are justified. In this limit, the relation $g_* \propto N^2$ is a generic prediction of our analysis. 

One interesting conclusion from our result Eq. (\ref{g*result1}) is that a  small enough value of $c_s$ may
help to relax the observational bound on $R$. For example, imposing the observational constraint $| g_*| \lesssim 10^{-2}$,  from the conventional formula Eq. (\ref{deltaP-ratio2})
one obtains the tight bound $R \lesssim 10^{-9}$. However, using our more general result  Eq. (\ref{g*result1}) this bound
relaxes to $R \sim \frac{\overline M_1}{M_P^2 \epsilon} \lesssim 10^{-9} c_s^{-5}$. Of course, this is based on the assumption that $c_s$ does not appear strongly in background parameters such as $\overline M_1$. It would be interesting to perform the analysis in a particular model of k-inflation to verify the above conclusion. 

\subsection{The case  $M_2, M_4 \neq 0$}

Now we extend the previous analysis to case in which $M_2$ and $M_4$ are non-zero. These  are
the coefficients of $\delta \left( G_{\alpha \beta} G^{\alpha \beta} \right)^2$  and 
$\delta \left( { G}_{\alpha \beta} { \tilde G}^{\alpha \beta} \right)^2$ in our starting unitary gauge action (\ref{action-UG}) which also appear in the quadratic action of transverse modes in Eq. (\ref{action-X}). 
Compared to Maxwell theory, these
are the terms  $\delta \left( F_{\alpha \beta} F^{\alpha \beta} \right)^2$ and $\delta \left( { F}_{\alpha \beta} { \tilde F}^{\alpha \beta} \right)^2$ which are the fourth orders in derivatives and are non-renormalizable. In quantum electrodynamics these interactions represent the photon-photon scattering and is known as the Euler-Heisenberg Lagrangian.  In the spirit of EFT these terms are irrelevant in low energy processes compared to terms coming with from $M_1$ and $\lambda_1$. In this view, $M_2, M_4  \lesssim  M_1 E_c^{-2}$ in which $E_c$ is the cutoff of the EFT. At an energy scale $E \ll E_c$ in which EFT is applicable, the contribution of the term containing $M_2$ and $M_4$ 
in the action  compared to
the leading term containing $M_1$ is approximately  given by $\frac{M_2 E^4}{M_1 E^2} \sim (\frac{E}{E_c})^2 \ll 1$.
Therefore,  in our analysis below,  the effects of $M_2$ and $M_4$ should be viewed as small sub-leading corrections compared to those of $M_1$ and $\lambda_1$. 

Unlike $M_1$ and $\lambda_1$ the  interactions  $M_2$ and $M_4$ do not show up explicitly in the interaction Lagrangian and  in exchange vertices in  Eqs. (\ref{L1}),   (\ref{L2}) and  (\ref{L3}).   However,   as can be seen from the quadratic action Eq. (\ref{action-X}),  their presence  affects the wave functions of $\delta X_T$  so their presences are felt via the corrections in  $\delta X_{T 1} $ in  $L_i$ in Eqs. (\ref{L1}),   (\ref{L2}) and  (\ref{L3}).

To calculate the corrections from $M_2$ and $M_4$ in   $\delta X_{T 1} $ it is much easier to work in linear polarization bases given in Eq. (\ref{linear-base}). Expanding  $\delta X_T $ in linear base as 
\ba
\delta X_{T i}= \delta X_T^{(1)} \epsilon^{(1)}_i +  \delta X_T^{(2)} \epsilon^{(2)}_i
\ea
yields 
\ba
\delta X_{T 1} =  - \delta X_T^{(1)}  \sin \theta , \quad 
\delta X_{T 2} =   \delta X_T^{(1)}  \cos \theta , \quad 
\delta X_{T 3} =   \delta X_T^{(2)}  \, .
\ea
Consequently, for the corresponding terms in Eq. (\ref{action-X}) we easily obtain 
\ba
\left(\partial_i \delta X_{T1}\right)^2 =   \left(\partial_i \delta X^{(1)}_{T}\right)^2  \sin ^2 \theta
\ea
and
\ba
\label{X2-cond}
\epsilon^{1 ij} \delta X_{T i, j} ={i  k} \delta X_T^{(2)} \sin \theta 
\ea
The above relation indicates that the mode $\delta X_T^{(2)}$ does not affect the power spectrum to leading order.
This is because $\pi^0$ couples only to $\delta X_{T 1} =  - \delta X_T^{(1)}  \sin \theta$ in interaction Lagrangians $L_1, L_2$ and $L_3$. In addition, from Eq. (\ref{X2-cond}) we find that 
the term containing $M_4$ in action (\ref{action-X}) contains only $\delta X_T^{(2)}$ which does not couple 
to $\pi^0$. Therefore, the effects of $M_4$ to  anisotropy power spectrum
 can be ignored to leading orders. 

Now, working only with the relevant component $\delta X_T^{(1)}$, the  action (\ref{action-X}) yields
\ba
S_2^{X_T^ {(1)} } &=& \int d^4 x \sqrt{-g} 
\Big[  \frac{1}{2} \big( \delta \dot{X}^{(1)c}_{T}\big)^2 -\frac{1}{2 a^2}
\Big(  1+  \frac{8\overline{M}_2H^2(2+n)^2 \sin^2 \theta }{\overline{M}_1-8\overline{M}_2H^2(2+n)^2}
\Big) \big( \delta X^{(1)c}_{T,j}\big)^2
\Big] 
 \label{action-X1}
\ea 
in which the relation between the normalized field ${X}^{(1)c}_{T}$ and ${X}^{(1)}_{T}$ is given as in
Eq. (\ref{normalization}).  The above action suggests that the speed of propagation for ${X}^{(1)c}_{T}$ 
is different than unity, given by (neglecting O( $\overline M_2^2$) )
\begin{equation}
\label{cv-eq}
c_v^2 \simeq1+\frac{{8} \overline{M}_2H^2(2+n)^2}{\overline{M}_1}\sin ^2\theta .
\end{equation}
There are two interesting conclusions here. First,  depending on the sign of $M_2$, the speed of propagation of
${X}^{(1)c}_{T}$ can be super-luminal or sub-luminal. Second, this speed  also depends on the direction of 
mode propagation,  given by the angle $\theta(\widehat {\bf k})$. Through $c_v$, these non-trivial effects also show up in the power spectrum anisotropy which may be interpreted as birefringence-like phenomena.

The wave function of the normalized field is
\ba
\delta X_{T}^{(1)c}=\frac{i H }{\sqrt{2(k c_v)^3}}(1 + i kc_v\tau) e^{-ikc_v\tau}
\ea
After taking into account the normalization relation between ${\delta X}^{(1)c}_{T}$ and ${\delta X}^{(1)}_{T}$ given in Eq. (\ref{normalization}),  for  ${\delta X}^{(1)}_{T}$ which appears in the interaction Lagrangians 
we obtain
\ba
\delta X_{T}^{(1) }=\frac{-i  \sin \theta  }{ H \tau^2 \sqrt{2 \overline M_1 (k c_v)^3}} 
\frac{(1 + i kc_v\tau) e^{-ikc_v\tau}}{\sqrt{1- \frac{c_v^2-1}{\sin^2 \theta }}}  \, .
\label{M_2wave}
\ea
The interactions are given as before by Eqs. (\ref{L1}),  (\ref{L2}) and (\ref{L3}) with ${\delta X}^{(1)}_{T}$ given above. 
 Performing the in-in integrals 
as before, the corrections in power spectrum is obtained to be
\ba
\frac{\delta P}{\delta P_{\pi^0}} &=& \frac{8 H^2 \overline M_1 c_s^2}{9 | \alpha_0| }
 \frac{(n+2)^3  (1+ \frac{6 \overline \lambda_1}{\overline M_1})  }{1- \frac{c_v^2-1}{\sin^2 \theta } } 
\Bigg[  \frac{c_s^3}{c_v^3} (n+2) (1+ \frac{6 \overline \lambda_1}{\overline M_1}) + (n- 1)  \Bigg] \, 
N^2 \sin^2 \theta  \nonumber\\
&=&  \frac{72  \overline M_1}{ \epsilon M_P^2 }
 \frac{  (1+ \frac{6 \overline \lambda_1}{\overline M_1})^2    c_s^5}{c_v^3( 1- \frac{c_v^2-1}{\sin^2 \theta })  }  \, 
N^2 \sin^2 \theta
\ea 
in which the second line  is obtained allowing $n=1$. 
Note in particular that when $M_2=0$ and $c_v=1$, the above result reduces to Eq. (\ref{deltaP-ratio1}) as expected. 

Now let us apply the above result to the conventional model of anisotropic inflation based on Maxwell theory 
as summarized below Eq. (\ref{g*result1}) in previous sub-section. We obtain 
\ba
\frac{\delta P}{\delta P_{\pi^0}}=24 I N^2  \sin^2 \theta\,  \Big[ 1- \frac{{36} H^2 \overline M_2}{\overline M_1} ( 1- 3 \cos^2 \theta) \Big] \, .
\ea
We see that the presence of the non-renormalizable term $M_2$ modifies the shape of anisotropy. We have both 
$\ell=2$ and $\ell=4$ harmonics  for power anisotropy.  Also note that while both polarization $\delta X^{(1)}_T$ and
$\delta X^{(2)}_T$ contribute into leading statistical anisotropy, but it is only  $\delta X^{(1)}_T  $ which contributes into
the sub-leading corrections containing $\overline M_2$.

\subsection{The  case $M_3 \neq 0$}

Now we go back to renormalizable models and assume $M_2 = M_4=0$, but allow for a non-zero coupling
$M_3$ which is the coupling of the interaction $G_{\alpha \beta } {\tilde G}^{\alpha \beta}$. In terms of Maxwell theory this corresponds to the interaction $F_{\mu \nu} {\tilde F}^{\mu \nu}$.  It is well known that this interaction breaks the parity. Usually the coupling to this interaction is controlled by the vev of a pseudo scalar known as the axion. The phenomenology of this interaction has been extensively studied in \cite{Dimopoulos:2012av, Namba:2015gja, Bartolo:2015dga, Bartolo:2014hwa, Caprini:2014mja}. Our  analysis below will be somewhat similar to analysis 
performed in \cite{Caprini:2014mja, Bartolo:2015dga}.  

We note that, like the situation involving $M_2$ and $M_4$, the coupling $M_3$ affects the free wave function of $\delta X_{T}$ fields so its presence change the anisotropic power spectrum through the modification in  $\delta X_{T 1}$ in interaction Lagrangians $L_i$.  Therefore, similar to the case with $M_2 $ and $M_4$, our job
is to calculate the corrections in $\delta X_{T 1}$ wave function in the presence of $M_3$. 

With $M_2=M_3=0$, the quadratic action (\ref{action-X}) reduces to
\begin{equation}
\label{SX3}
S^X=\int d^4x \sqrt{-g}\Big[ \frac{1}{2}\big( \delta \dot{X}_i^c\big)^2 -\frac{1}{2 a^2}\big(\delta X_{i,j}^c\big)^2 
+ \frac{2 M_3 H a^4}{\overline M_1} \Big(n-\frac{3}{2}-\frac{\dot{M}_3}{2M_3H}\Big) \epsilon ^{ijk} \delta X^{(c)}_{Ti}\delta X^{(c)}_{Tj,k}\Big]
\end{equation}
in which $ \delta X^{(c)}_{T i}=\sqrt{M_1}a^{-1} \delta X_{Ti}$. 

To proceed further we need to find the  time variation of $M_3$. This term does not appear in the background since it gives rise to magnetic field which is zero for our choice of background containing only the electric field.  Therefore, the scaling of this term with time is free. However, if it scales differently form spatial gradient part  then 
things become non-trivial  from competition of these two terms during inflation. One intuitive argument to 
set the scaling of $M_3$ with time is to demand that the equation for the free wave function in Fourier space to
depend only on the combination  $k/a$. This is motivated from the fact that the physical wave number is $k/a$.
For example, the usual gradient term in  action (\ref{SX3}) yields $k^2/a^2$. Demanding that only
the combination $k/a$ appears for the term containing $M_3$ requires that   
$M_3 a^4 \propto k/a$ so we conclude 
\begin{equation}
\label{M3-scale}
M_3=\overline{M}_3 a^{-5}.
\end{equation}
We mention   that the above argument may provide a natural expectation for the scaling of $M_3$ as given
in Eq. (\ref{M3-scale}) but it does not seem exhaustive. As a result,  in principle,  one may allow for different time scaling than used in Eq. (\ref{M3-scale}). 

To solve the free wave function in the presence of $M_3$, this time  it is more convenient to switch to the helicity (circular) base given in Eq. (\ref{circular-base}) in which 
\begin{equation}
\delta X^{c}_{T i}(k)=\sum _{s=\pm} \delta X^{(s)}_T (k)\epsilon ^s_i(k) \, .
\end{equation}
Using the relation 
\ba
\epsilon^{m j l} \delta X^{c}_{T j, l} =  -k \sum_{s=\pm}  s \delta X^{(s)}_T  \epsilon^s_m(k) \, ,
\ea
the equation of motion for the free wave function is obtained to be 
\begin{equation}
\delta \ddot{X}^{(s)}_T+3H\delta \dot{X}^{(s)}_T+\frac{k^2}{a^2}\delta X^{(s)}_T+4(n+1)s \frac{k}{a}\frac{\overline{M}_3H}{\overline{M}_1}\delta X^{(s)}_T=0.
\end{equation}
As demanded, the coefficients in the  above equation depend on the combination $\frac{k}{a}$. 
Now going to conformal time and defining  $\delta X^{(s)}=-H\tau \delta V^{(s)}$ we obtain,
\begin{equation}
\delta V^{(s)\prime\prime}+\left(k^2+\frac{2sk}{\tau}\xi -\frac{2}{\tau ^2}\right) \delta V^{(s)} =0, 
\label{e.o.mM-1}
\end{equation} 
where we have defined 
\begin{equation}
\xi = -2(n+1) \frac{\overline{M_3}}{\overline{M}_1}.
\end{equation}
As expected, this equation has the same form in the model studied in  \cite{Caprini:2014mja, Bartolo:2015dga} so
our argument from here will be mostly similar to those of  \cite{Caprini:2014mja, Bartolo:2015dga}. 

If $|\xi| <<1$ then the effect of term containing $M_3$ will be suppressed cosmologically, and so, we may consider opposite limit in which $|\xi|>>1$. With this assumption the phenomenology originating from equation \eqref{e.o.mM-1} is very interesting. Note that for scales deep inside the horizon only the first term in the bracket in Eq. (\ref{e.o.mM-1})  is important so both of the polarization are in Minkowski vacuum as expected. On the other hand, in the regime   $0\ll  |k\tau| \ll  |\xi |$ the second term dominates while its sign depends on polarizations through the pre-factor $s$ and the sign of $\xi$. For the moment let us assume  that $\xi >0$ so from
Eq. (\ref{e.o.mM-1}) we see that  only the positive helicity, $s=+$,  is  amplified so at the end of inflation $\delta X^{(s)}$ is highly polarized with positive helicity. Inversely, if $\xi <0$, then the negative  helicity is amplified and $\delta X^{(s)}$ becomes a pure negative helicity at the end of inflation. However, the overall amplitude of these polarizations are the same on super-horizon scales and as their couplings to $\pi ^0$ are also the same,  the final result will not change. As a result, without loss of generality,   we may simply take $\xi>0$. 

The general solution of  Eq. (\ref{e.o.mM-1}) is presented in \cite{Caprini:2014mja} which on super-horizon scales, $k \tau \rightarrow 0$,  simplifies to 
\begin{equation}
\delta V^+=\frac{e^{\pi \xi}}{\xi ^{3/2}}\frac{(-\tau)^{-1}}{2\sqrt{\pi k^3}}, \qquad k\tau \rightarrow 0,
\end{equation}
which leads to,
\begin{equation}
\delta X_{T1}=\frac{-i\sin \theta}{\sqrt{2 \overline{M}_1k^3}H\tau ^2} \frac{e^{\pi \xi}}{2\sqrt{\pi \xi ^3}} \qquad k\tau \rightarrow 0. \label{M_1wave1}
\end{equation}
The in-in integrals are easy to calculate noting that  the main contribution to the in-in integrals comes from super-horizon scales. Now comparing the wave function in  \eqref{M_1wave1} with the wave function in \eqref{M_2wave} towards  the end of inflation, it is easy to see that the only difference here is that $c_v=1$ while  the power of $\delta X_{T1}$ will be amplified with the additional factor  $(\frac{e^{\pi |\xi|}}{2\sqrt{\pi |\xi| ^3}})^2$.  Therefore, we obtain 
 \ba
g_*&=& \frac{8 H^2 \overline M_1 c_s^2}{9 | \alpha_0|} \left(\frac{e^{2 \pi |\xi|}}{4 {\pi |\xi| ^3}}\right)
 \left[(n+2)^3 N^2 (1+ \frac{6 \overline \lambda_1}{\overline M_1}) \sin^2 \theta \right]
\Big[  c_s^3 (n+2) (1+ \frac{6 \overline \lambda_1}{\overline M_1}) +(n- 1)  \Big] \nonumber\\
&=&  \frac{72 H^2 \overline M_1 c_s^5}{ | \alpha_0|} \left(\frac{e^{2 \pi |\xi|}}{4 {\pi |\xi| ^3}}\right)
  \left(1+ \frac{6 \overline \lambda_1}{\overline M_1} \right)^2  N^2  \sin^2 \theta  \, ,
\ea
in which the final result is obtained setting $n=1$. 

In particular, for the model studied in \cite{Caprini:2014mja, Bartolo:2015dga} with $ c_s=1, \lambda_1=0, 
\overline M_1 = \frac{I \epsilon}{3} M_P^2$ and $\alpha_0 =- \epsilon H^2 M_P^2$ we obtain
 \ba
g_*=  24 I  \left(\frac{e^{2 \pi |\xi|}}{4 {\pi |\xi| ^3}}\right)   N^2 \sin^2 \theta 
\ea
in  agreements with the results of \cite{Caprini:2014mja, Bartolo:2015dga}.


\section{Summary and Discussions}
\label{Summary sec.}

As argued before, EFT of inflation is a powerful tool to study inflation model-independently.
In particular,  EFT approach is very helpful to classify different models of inflation based on their
predictions for power spectrum and bispectrum.  So far most of the EFT studies were based in inflation in FRW setup involving scalar fields. In these setups 
one chooses a  space-time foliation which sets the scalar field fluctuations to zero.
Consequently, all perturbations are transferred into metric perturbations. However, the system enjoys  the remnant 
three-dimensional  diffeomorphism invariance $x^i \rightarrow x^i + \xi^i (x^\nu)$. Having presented the most general action in unitary gauge which  respects the remnant symmetry,  one obtains all interactions
after restoring  the Goldstone boson $\pi$ associated with the time diffeomorphism breaking . 

Our goal in this study was to extend the EFT approach to the models of anisotropic inflation in which a  
background gauge field, in the form of an electric field, contributes to the inflationary dynamics, for
relevant works but in different setups see \cite{Cannone:2015rra, Hidaka:2014fra, Lin:2015cqa}. 
The background is intrinsically anisotropic in the form of Bianchi I universe. To simplify the analysis we work in the decoupling limit where
the gravitational back-reactions are negligible on dynamics of $\delta \phi$ and $\delta A_\mu$ perturbations. In particular, within this assumption, one can approximate the Bianchi I background by the usual FRW metric and take all three scale
factors to be the same as far as the gauge field perturbations are concerned. Physically, this means that the leading contributions to  statistical anisotropies are sourced by matter perturbations. This was specifically demonstrated in the simple model of anisotropic inflation in \cite{Emami:2013bk}. 

The important task in our analysis was to understand the nature of the underlying symmetry and to 
read off the physical degrees of freedom. These  are the necessary steps to define the unitary gauge and  to present the starting general action invariant under the remnant symmetry. As in single field model of inflation, we can still use 
inflaton as the proper clock to define our time foliation. However, the situations with gauge field is more non-trivial. This is mainly because we have to enforce the $U(1)$ gauge symmetry on gauge field perturbations. Putting specifically, even if we start with $\delta A^\mu=0$, there is always a $U(1)$ gauge transformation which can restore $\delta A^\mu$. Upon 
taking care of both coordinate diffeomorphism and the $U(1)$ invariance we have identified the remnant symmetry of
the system as given in Eq. (\ref{UG-symmetry}).  Obviously this symmetry is smaller than the remnant symmetry in single 
field model with no gauge field. However, thanks to the crucial roles of the $U(1)$ gauge symmetry,  
this remnant symmetry is still large enough to prevent the appearance of pathologies such as ghost or tachyon. 
Indeed we have checked that if one does not reinforce the gauge symmetry, i.e. take $A^\mu$ as a mere 4-vector, the remnant symmetry is smaller than Eq. (\ref{UG-symmetry}) and many new terms pop up in the unitary gauge action. The situation may get out of control as some of the new terms may have ghosts and other unwanted pathologies. This seems an interesting question and we would like to come back to this question elsewhere. 

Having presented the proper unitary gauge and the corresponding remnant symmetry, we have identified the 
building blocks to present the invariant action as given in Eq. (\ref{action-UG}).  As we have seen the coupling $M_1$
represents the known models of anisotropic inflation based on Maxwell theory. Interestingly, the couplings 
$\lambda_1, \lambda_2, M_2, M_3$ and $M_4$ represent new types of interaction. Also the parameter 
$c_0$ measures the sound speed of curvature perturbations. Upon performing the so-called Stueckelberg trick, we restore 
the Goldstone bosons.  In total 
we have three Goldstone bosons,  $\pi^0$ and $\delta X_{T i} = a^2 \partial_1 \pi_T^i$, in which $\pi^0$
is associated with breaking the time diffeomorphism, representing the inflaton perturbations. The other two Goldstone bosons represent the two transverse polarization degrees of freedom of gauge field fluctuations.

After presenting the wave function of the free theory and the leading interactions, we have calculated the anisotropy corrections to curvature perturbation power spectrum for various couplings. As expected, we have recovered the known results for power anisotropies in known models of anisotropic inflation. In addition, we have shown that the sound speed
$c_s$ and the coupling $\lambda_1$  can play non-trivial roles.  We have seen that the non-renormalizable term containing $M_2$ introduce a phenomena similar to birefringence in which the speed of gauge field
propagation $c_v$ depends on the direction of the propagating mode. In addition, the two polarization of gauge fields 
contribute asymmetrically in curvature perturbation anisotropies. Finally, we have seen that the coupling $M_3$ captures 
the parity violating model $F \tilde F$ as studied in the past literature.

We comment that here we have assumed that there is only one scalar field degree of freedom. If one considers the cases 
involving multiple light scalar fields then the number of Goldstone bosons will be different. For example, if 
one starts with the charged $U(1)$ setup in which the gauge field is charged under a complex scalar field, as studied
in \cite{Emami1, Emami:2013bk},  then one expects 
to have four Goldstone bosons. Assuming that the potential is a function of the radial part of the complex scalar 
field, then the  additional Goldstone boson represents the axial part of the complex scalar field. Upon Higgs symmetry breaking, the scalar's axial degree of freedom  is eaten by the gauge field, creating its longitudinal degree of freedom.
It will be an interesting exercise to present the EFT description of these symmetry breaking scenarios.  

There are few directions which we would like to pursue in future works. One natural question is the bispectrum analysis.
As is well-known, the EFT approach is specially powerful in non-Gaussianity analysis. 
Therefore one expects that our EFT approach will be very rich in understanding the generic features of non-Gaussianity in
models of anisotropic inflation. The bispectrum and trispectrum  for simple models of anisotropic inflation
were studied in   \cite{Abolhasani:2013zya,  Bartolo:2012sd, Shiraishi:2013vja}. We would like to study  model-independently  the implications of our EFT approach for non-Gaussianity.
Another question is the role of gravitational waves. As is well-known, in models of anisotropic inflation 
there will be mixing between the curvature perturbations and tensor perturbations $h_{ij}$ yielding a non-zero cross-correlation $\langle \zeta h_{ij} \rangle$.
This effect was studied in \cite{Chen:2014eua, Bartolo:2015dga}. We expect our general EFT approach to go beyond these 
analysis and yield more non-trivial results for CMB $TT, TB$ or $EB$ cross-correlations.  Also the question of statistical anisotropies  beyond the lore of anisotropic inflation in which there is no gauge field and the anisotropies are generated by
a generic four vector is another question of interest. In these setups the remnant symmetries are even smaller than the model with $U(1)$ fields and many interactions are allowed. Finding a healthy theory within this setup and looking for
their predictions for statistical anisotropies is an interesting question which deserves further investigations.

\vspace{0.7cm}
{\bf Acknowledgement:} We  thank W. Goldberger,  J. Gong,  E. Komatsu,  M. Mirbabayi,    L. Senatore, J. Soda, 
G. Tasinato  and M. Yamaguchi for useful discussions and correspondences. 
H. F. would like to thank Munich Institute for Astro- and Particle Physics (MIAPP) 
for the hospitality  during the workshop `` Cosmology after Planck"  where this work was in progress.
R.E is supported by Grant HKUST4/CRF/13G issued by the Research Grants
Council (RGC) of Hong Kong.

{}


\begin{thebibliography}{}


\bibitem{Ade:2013lta}
  P.~A.~R.~Ade {\it et al.}  [Planck Collaboration],
  arXiv:1303.5076 [astro-ph.CO].

\bibitem{Ade:2013uln}
  P.~A.~R.~Ade {\it et al.}  [Planck Collaboration],
  arXiv:1303.5082 [astro-ph.CO].
  
  
  
\bibitem{Cheung:2007st} 
  C.~Cheung, P.~Creminelli, A.~L.~Fitzpatrick, J.~Kaplan and L.~Senatore,
  JHEP {\bf 0803}, 014 (2008)
  [arXiv:0709.0293 [hep-th]].

\bibitem{Manohar:1996cq} 
  A.~V.~Manohar,
  Lect.\ Notes Phys.\  {\bf 479}, 311 (1997)
  [hep-ph/9606222].


\bibitem{Burgess:2007pt} 
  C.~P.~Burgess,
  Ann.\ Rev.\ Nucl.\ Part.\ Sci.\  {\bf 57}, 329 (2007)
  [hep-th/0701053].


\bibitem{Senatore:2010wk} 
  L.~Senatore and M.~Zaldarriaga,
  JHEP {\bf 1204}, 024 (2012)
  [arXiv:1009.2093 [hep-th]].
  


\bibitem{Emami:2015qjl} 
  R.~Emami,
  arXiv:1511.01683 [astro-ph.CO].
		
\bibitem{Watanabe:2009ct}
  M.~a.~Watanabe, S.~Kanno and J.~Soda,
  Phys.\ Rev.\ Lett.\  {\bf 102}, 191302 (2009)
  [arXiv:0902.2833 [hep-th]].
  
 \bibitem{Watanabe:2010fh}
  M.~a.~Watanabe, S.~Kanno and J.~Soda,
  Prog.\ Theor.\ Phys.\  {\bf 123}, 1041 (2010)
  [arXiv:1003.0056 [astro-ph.CO]].

 \bibitem{Soda1} 
J.~Ohashi, J.~Soda and S.~Tsujikawa,
JCAP {\bf 1312}, 009 (2013)
[arXiv:1308.4488 [astro-ph.CO], arXiv:1308.4488].

J.~Ohashi, J.~Soda and S.~Tsujikawa,
Phys.\ Rev.\ D {\bf 88}, 103517 (2013)
[arXiv:1310.3053 [hep-th]].

J.~Ohashi, J.~Soda and S.~Tsujikawa,
Phys.\ Rev.\ D {\bf 87}, 083520 (2013)
[arXiv:1303.7340 [astro-ph.CO]].

S.~Kanno, J.~Soda, M.~-a.~Watanabe,
JCAP {\bf 1012}, 024 (2010).
[arXiv:1010.5307 [hep-th]].

K.~Murata, J.~Soda,
JCAP {\bf 1106}, 037 (2011).
[arXiv:1103.6164 [hep-th]].

S.~Yokoyama and J.~Soda,
JCAP {\bf 0808}, 005 (2008);

M.~-a.~Watanabe, S.~Kanno and J.~Soda,
Mon.\ Not.\ Roy.\ Astron.\ Soc.\  {\bf 412}, L83 (2011)
[arXiv:1011.3604 [astro-ph.CO]].

K.~Yamamoto, M.~-a.~Watanabe and J.~Soda,
Class.\ Quant.\ Grav.\  {\bf 29}, 145008 (2012)
[arXiv:1201.5309 [hep-th]].

  A.~Ito and J.~Soda,
  arXiv:1506.02450 [hep-th].


\bibitem{Emami1}
R.~Emami, H.~Firouzjahi, S.~M.~Sadegh Movahed, M.~Zarei,
JCAP {\bf 1102 } (2011)  005.
[arXiv:1010.5495 [astro-ph.CO]].

R.~Emami and H.~Firouzjahi,
JCAP {\bf 1201}, 022 (2012)
[arXiv:1111.1919 [astro-ph.CO]].


  S.~Baghram, M.~H.~Namjoo and H.~Firouzjahi,
  JCAP {\bf 1308}, 048 (2013)
  [arXiv:1303.4368 [astro-ph.CO]].



  R.~Emami and H.~Firouzjahi,
  JCAP {\bf 1510}, no. 10, 043 (2015)
  [arXiv:1506.00958 [astro-ph.CO]].



\bibitem{Emami:2013bk}
R.~Emami and H.~Firouzjahi,
JCAP {\bf 1310}, 041 (2013)
[arXiv:1301.1219 [hep-th]].

\bibitem{Abolhasani:2013zya}
A.~A.~Abolhasani, R.~Emami, J.~T.~Firouzjaee and H.~Firouzjahi,
JCAP {\bf 1308}, 016 (2013)
[arXiv:1302.6986 [astro-ph.CO]].

\bibitem{Abolhasani:2013bpa}
A.~A.~Abolhasani, R.~Emami and H.~Firouzjahi,
arXiv:1311.0493 [hep-th].

 \bibitem{Chen:2014eua}
 X.~Chen, R.~Emami, H.~Firouzjahi and Y.~Wang,
 arXiv:1404.4083 [astro-ph.CO].

\bibitem{Bartolo:2012sd}
  N.~Bartolo, S.~Matarrese, M.~Peloso and A.~Ricciardone,
  Phys.\ Rev.\ D {\bf 87}, 023504 (2013)
  [arXiv:1210.3257 [astro-ph.CO]].
  
  \bibitem{Shiraishi:2013vja}
M.~Shiraishi, E.~Komatsu, M.~Peloso and N.~Barnaby,
JCAP {\bf 1305}, 002 (2013)
[arXiv:1302.3056 [astro-ph.CO]].
  
  \bibitem{Shiraishi:2013oqa}
M.~Shiraishi, E.~Komatsu and M.~Peloso,
arXiv:1312.5221 [astro-ph.CO].



\bibitem{various}

  K.~Dimopoulos, M.~Karciauskas, D.~H.~Lyth and Y.~Rodriguez,
  JCAP {\bf 0905}, 013 (2009)
  [arXiv:0809.1055 [astro-ph]].


  A.~E.~Gumrukcuoglu, B.~Himmetoglu, M.~Peloso,
  Phys.\ Rev.\  {\bf D81}, 063528 (2010).
  [arXiv:1001.4088 [astro-ph.CO]].

T.~R.~Dulaney, M.~I.~Gresham,
Phys.\ Rev.\  {\bf D81}, 103532 (2010).
[arXiv:1001.2301 [astro-ph.CO]].


K.~Yamamoto,
Phys.\ Rev.\ D {\bf 85}, 123504 (2012)
[arXiv:1203.1071 [astro-ph.CO]].
  
\bibitem{Funakoshi:2012ym} 
  H.~Funakoshi and K.~Yamamoto,
  Class.\ Quant.\ Grav.\  {\bf 30}, 135002 (2013)
  [arXiv:1212.2615 [astro-ph.CO]].

  
T.~Fujita and S.~Yokoyama,
JCAP {\bf 1309}, 009 (2013)
[arXiv:1306.2992 [astro-ph.CO]].


S.~R.~Ramazanov and G.~Rubtsov,
Phys.\ Rev.\ D {\bf 89}, 043517 (2014)
[arXiv:1311.3272 [astro-ph.CO]].


  S.~Nurmi and M.~S.~Sloth,
  JCAP {\bf 1407}, 012 (2014)
  [arXiv:1312.4946 [astro-ph.CO]].


  R.~K.~Jain and M.~S.~Sloth,
  JCAP {\bf 1302}, 003 (2013)
  [arXiv:1210.3461 [astro-ph.CO]].


  F.~R.~Urban,
  Phys.\ Rev.\ D {\bf 88}, 063525 (2013)
  [arXiv:1307.5215 [astro-ph.CO]].


M.~Thorsrud, D.~F.~Mota and S.~Hervik,
JHEP {\bf 1210}, 066 (2012)
[arXiv:1205.6261 [hep-th]].

S.~Bhowmick and S.~Mukherji,
Mod.\ Phys.\ Lett.\ A {\bf 27}, 1250009 (2012)
[arXiv:1105.4455 [hep-th]].

  S.~Hervik, D.~F.~Mota and M.~Thorsrud,
  JHEP {\bf 1111}, 146 (2011)
  [arXiv:1109.3456 [gr-qc]].



C.~G.~Boehmer, D.~F.~Mota,
Phys.\ Lett.\  {\bf B663}, 168-171 (2008).
[arXiv:0710.2003 [astro-ph]].

T.~S.~Koivisto, D.~F.~Mota,
JCAP {\bf 0808}, 021 (2008).
[arXiv:0805.4229 [astro-ph]].

J.~P.~Beltran Almeida, Y.~Rodriguez and C.~A.~Valenzuela-Toledo,
vector fields,''
Mod.\ Phys.\ Lett.\ A {\bf 28} (2013) 1350012
[arXiv:1112.6149 [astro-ph.CO]].

  Y.~Rodriguez, J.~P.~Beltran Almeida and C.~A.~Valenzuela-Toledo,
  JCAP {\bf 1304}, 039 (2013)
  [arXiv:1301.5843 [astro-ph.CO]].


D.~H.~Lyth and M.~Karciauskas,
JCAP {\bf 1305}, 011 (2013)
[arXiv:1302.7304 [astro-ph.CO]].

Tuan Q. Do and W. F. Kao,
Phys. Rev. D 84, 123009.\\
Tuan Q. Do, W. F. Kao, and Ing-Chen Lin,
``Anisotropic power-law inflation for a two scalar fields model,'' Phys. Rev. D 83, 123002.

\bibitem{Naruko:2014bxa} 
A.~Naruko, E.~Komatsu and M.~Yamaguchi,
  JCAP {\bf 1504}, no. 04, 045 (2015)
  [arXiv:1411.5489 [astro-ph.CO]].


\bibitem{Kim:2013gka}
J.~Kim and E.~Komatsu,
Phys.\ Rev.\ D {\bf 88}, 101301 (2013)
[arXiv:1310.1605 [astro-ph.CO]].


\bibitem{various2} 

  N.~Bartolo, S.~Matarrese, M.~Peloso and A.~Ricciardone,
  JCAP {\bf 1308}, 022 (2013)
  [arXiv:1306.4160 [astro-ph.CO]].

  N.~Bartolo, M.~Peloso, A.~Ricciardone and C.~Unal,
  JCAP {\bf 1411}, no. 11, 009 (2014)
  [arXiv:1407.8053 [astro-ph.CO]].

  M.~Akhshik, R.~Emami, H.~Firouzjahi and Y.~Wang,
  JCAP {\bf 1409}, 012 (2014)
  [arXiv:1405.4179 [astro-ph.CO]].

  M.~Akhshik,
  JCAP {\bf 1505}, no. 05, 043 (2015)
  [arXiv:1409.3004 [astro-ph.CO]].

  X.~Chen, R.~Emami, H.~Firouzjahi and Y.~Wang,
  JCAP {\bf 1504}, no. 04, 021 (2015)
  [arXiv:1408.2096 [astro-ph.CO]].


  X.~Li, S.~Wang and Z.~Chang,
  Eur.\ Phys.\ J.\ C {\bf 75}, no. 6, 260 (2015)
  [arXiv:1502.02256 [gr-qc]].

  C.~Pitrou, T.~S.~Pereira and J.~P.~Uzan,
  JCAP {\bf 0804}, 004 (2008)
  [arXiv:0801.3596 [astro-ph]].


  G.~Esposito-Farese, C.~Pitrou and J.~P.~Uzan,
  Phys.\ Rev.\ D {\bf 81}, 063519 (2010)
  [arXiv:0912.0481 [gr-qc]].



\bibitem{Ackerman:2007nb} 
  L.~Ackerman, S.~M.~Carroll and M.~B.~Wise,
  Phys.\ Rev.\ D {\bf 75}, 083502 (2007)
  [Phys.\ Rev.\ D {\bf 80}, 069901 (2009)]
  [astro-ph/0701357].

\bibitem{Pullen:2007tu} 
  A.~R.~Pullen and M.~Kamionkowski,
  Phys.\ Rev.\ D {\bf 76}, 103529 (2007)
  [arXiv:0709.1144 [astro-ph]].



\bibitem{ArmendarizPicon:1999rj} 
  C.~Armendariz-Picon, T.~Damour and V.~F.~Mukhanov,
  Phys.\ Lett.\ B {\bf 458}, 209 (1999)
  [hep-th/9904075].

\bibitem{Garriga:1999vw} 
  J.~Garriga and V.~F.~Mukhanov,
  Phys.\ Lett.\ B {\bf 458}, 219 (1999)
  [hep-th/9904176].



\bibitem{Alishahiha:2004eh} 
  M.~Alishahiha, E.~Silverstein and D.~Tong,
  Phys.\ Rev.\ D {\bf 70}, 123505 (2004)
  [hep-th/0404084].

\bibitem{Dimopoulos:2011pe} 
  K.~Dimopoulos, D.~Wills and I.~Zavala,
  Nucl.\ Phys.\ B {\bf 868}, 120 (2013)
  [arXiv:1108.4424 [hep-th]].


\bibitem{Weinberg:2005vy}
  S.~Weinberg,
  Phys.\ Rev.\ D {\bf 72}, 043514 (2005)
  [hep-th/0506236].

\bibitem{Chen:2010xka}
  X.~Chen,
 Adv.\ Astron.\  {\bf 2010}, 638979 (2010)  [arXiv:1002.1416 [astro-ph.CO]].  
   
  
\bibitem{Wang:2013eqj} 
  Y.~Wang,
  Commun.\ Theor.\ Phys.\  {\bf 62}, 109 (2014)
  [arXiv:1303.1523 [hep-th]].

  
  
\bibitem{Dimopoulos:2012av} 
  K.~Dimopoulos and M.~Karciauskas,
  JHEP {\bf 1206}, 040 (2012)
  [arXiv:1203.0230 [hep-ph]].
 
  
\bibitem{Namba:2015gja} 
  R.~Namba, M.~Peloso, M.~Shiraishi, L.~Sorbo and C.~Unal,
  arXiv:1509.07521 [astro-ph.CO].

\bibitem{Bartolo:2015dga} 
  N.~Bartolo, S.~Matarrese, M.~Peloso and M.~Shiraishi,
  JCAP {\bf 1507}, no. 07, 039 (2015)
  [arXiv:1505.02193 [astro-ph.CO]].

\bibitem{Bartolo:2014hwa} 
  N.~Bartolo, S.~Matarrese, M.~Peloso and M.~Shiraishi,
  JCAP {\bf 1501}, no. 01, 027 (2015)
  [arXiv:1411.2521 [astro-ph.CO]].

\bibitem{Caprini:2014mja} 
  C.~Caprini and L.~Sorbo,
  JCAP {\bf 1410}, no. 10, 056 (2014)
  [arXiv:1407.2809 [astro-ph.CO]].

\bibitem{Cannone:2015rra} 
  D.~Cannone, J.~O.~Gong and G.~Tasinato,
  JCAP {\bf 1508}, no. 08, 003 (2015)
  [arXiv:1505.05773 [hep-th]].

\bibitem{Hidaka:2014fra} 
  Y.~Hidaka, T.~Noumi and G.~Shiu,
  Phys.\ Rev.\ D {\bf 92}, no. 4, 045020 (2015)
  [arXiv:1412.5601 [hep-th]].


\bibitem{Lin:2015cqa} 
  C.~Lin and L.~Z.~Labun,
  arXiv:1501.07160 [hep-th].





\end{thebibliography}
\end{document}